\documentclass[a4paper,11pt,twoside]{article} 

\usepackage{amsmath,amsfonts,amssymb}
\usepackage{graphicx}
\usepackage{caption}
\usepackage[subrefformat=parens,labelformat=parens]{subcaption}
\usepackage[margin=2.5cm]{geometry}
\usepackage{pdflscape}
\usepackage{afterpage}
\usepackage{longtable}
\usepackage{enumitem}
\usepackage{multirow}
\usepackage{cite}
\usepackage{hhline}
\usepackage{lineno}
\usepackage[table]{xcolor}
\usepackage{array,arydshln}
\usepackage{authblk}
\usepackage{tikz}
\usetikzlibrary{positioning}
\usepackage[skins]{tcolorbox}
\usepackage[colorlinks=true, allcolors=black]{hyperref}
\setlist{topsep=0ex,itemsep=-1ex,partopsep=0ex,parsep=1ex}

\definecolor{highlightcolor}{RGB}{255,245,0}
\definecolor{highlightcolor2}{RGB}{0,200,0}

\newcommand{\rotatecol}[1]{\makebox[0.75em][l]{\rotatebox{90}{#1}}}
\newcolumntype{P}[1]{>{\raggedright\arraybackslash}m{#1}}
\newcolumntype{C}[1]{>{\centering\arraybackslash}m{#1}}

\graphicspath{{../convert_figs/}}

\title{Single-photon SPAD imagers in biophotonics:\\ Review and Outlook}

\author[1,C,*]{Claudio Bruschini}
\author[2,*]{Harald Homulle}
\author[1]{Ivan Michel Antolovic}
\author[1]{Samuel Burri}
\author[1]{Edoardo Charbon}

\affil[1]{AQUA, EPFL, Neuch\^atel, Switzerland}
\affil[2]{AQUA, TU Delft, Delft, The Netherlands}
\affil[C]{Corresponding author: claudio.bruschini@epfl.ch}
\affil[*]{These authors contributed equally to the work presented in this article}
\date{}

\begin{document}

\maketitle

\footnotetext[1]{Article submitted to \it{Light: Science and Applications}}

\begin{abstract}
Single-photon avalanche diode (SPAD) arrays are solid-state detectors offering imaging capabilities at the level of individual photons, with unparalleled photon counting and time-resolved performance. This fascinating technology has progressed at very high pace in the past 15~years, since its inception in standard CMOS technology in 2003. A host of architectures has been explored, ranging from simpler implementations, based solely on off-chip data processing, to progressively ``smarter" sensors including on-chip, or even pixel-level, timestamping and processing capabilities. As the technology matured, a range of biophotonics applications has been explored, including (endoscopic) FLIM, (multi-beam multiphoton) FLIM-FRET, SPIM-FCS, super-resolution microscopy, time-resolved Raman, NIROT, and PET. We will review some representative sensors and their corresponding applications, including the most relevant challenges faced by chip designers and end-users. Finally, we will provide an outlook on the future of this fascinating technology.
\end{abstract}

\section{Introduction}\label{sec:intro}

Individual single-photon avalanche diodes (SPADs) have long been the detector of choice when deep sub-nanosecond timing performance was required, due to their excellent single-photon detection and timestamping capability \cite{Zappa2007, Zappa2010, Charbon2014, Perenzoni2016II}. What did, however, really trigger the exploration and design of large digital SPAD imagers, potentially manufactured in volume at affordable prices, was the breakthrough implementation of the first SPADs in standard CMOS technology \cite{Rochas2003SPAD} (2003). This was soon followed by the first integrated SPAD array \cite{Rochas2003} and a host of architectures, ranging from simpler implementations of the early days, based solely on off-chip data processing, to progressively ``smarter" sensors including on-chip, or even pixel-level, timestamping and processing capabilities. Modular setups have been designed as well, either through the combination of SPAD arrays with FPGAs (``reconfigurable pixels"), or by means of very recent 3D developments. Furthermore, basically all implementations rely on FPGA-based host boards; this, combined with the natively digital SPAD data output, opens the door to real-time algorithmic implementations in close sensor proximity, such as FPGA-based autocorrelation and lifetime calculations.

As the SPAD technology matured, a range of applications have been explored in very diverse fields, such as consumer and robotics imaging, data and telecom security, advanced driver-assistance systems, and biophotonics, which is the main subject of this review. We will discuss in particular (endoscopic) FLIM, (multi-beam multiphoton) FLIM-FRET, SPIM-FCS, localization- and entangled photons-based super-resolution microscopy, time-resolved Raman, NIROT, and PET. It is, however, true that SPAD imagers are still mostly used in specialized research settings, apart from some notable non-imaging exceptions, such as SPAD arrays in the form of silicon photo-multipliers (SiPMs), which are readily available from a number of manufacturers. This is at first look surprising, given the aforementioned potential for unrivalled photon counting and time-resolved performance, but can be partly traced back to some performance parameters which still lag behind those of established CCDs and sCMOS imagers, such as quantum efficiency over the whole spectrum and fill factor, which are of importance for several light-starved applications. The pixel sizes are typically larger, limiting so far the manufacturing of megapixel arrays. On the technology side, the design of high performance, low noise SPADs is challenging; the same is true at system level for data handling, leading to important firmware development efforts. It therefore comes to no surprise that recent efforts have been focused, at the device level, on increasing the SPADs’ key figures of merit \cite{Bronzi2016} and improving the contact with foundries, to fully profit from possible process optimizations.

In the following sections, which are mostly dedicated to SPAD imagers in standard CMOS technologies, we will first dwell on the SPAD state-of-the-art, starting from individual devices, their key properties and the corresponding optimization trade-offs, which are strongly influenced by the target application. We will then focus on the impact of design choices on the overall sensor architecture and the most important challenges, moving up in a vertical fashion from the pixel level, considering basic circuitries and in-pixel options, to array architectures (1D vs. 2D), and the read-out, which is of particular importance for real-time implementations. A host of biophotonics applications will then be described, starting from FLIM and its different flavours, to end with more disruptive scenarios and sensor concepts such as quantum-based super-resolution microscopy and 3D stacking (the combination of a top sensor layer with a bottom control and processing layer), respectively.

The interested reader is encouraged to look at \cite{Charbon2014} for details of other applications of SPAD-based imagers, and \cite{Esposito2012, Henderson2014, Suhling2015} for a comparison to established devices as well as alternative CMOS imagers.

 \section{SPAD detectors and imagers}\label{sec:SPADs} 

\subsection{Single-photon avalanche diodes}

	\begin{figure}
		\includegraphics[width=\textwidth]{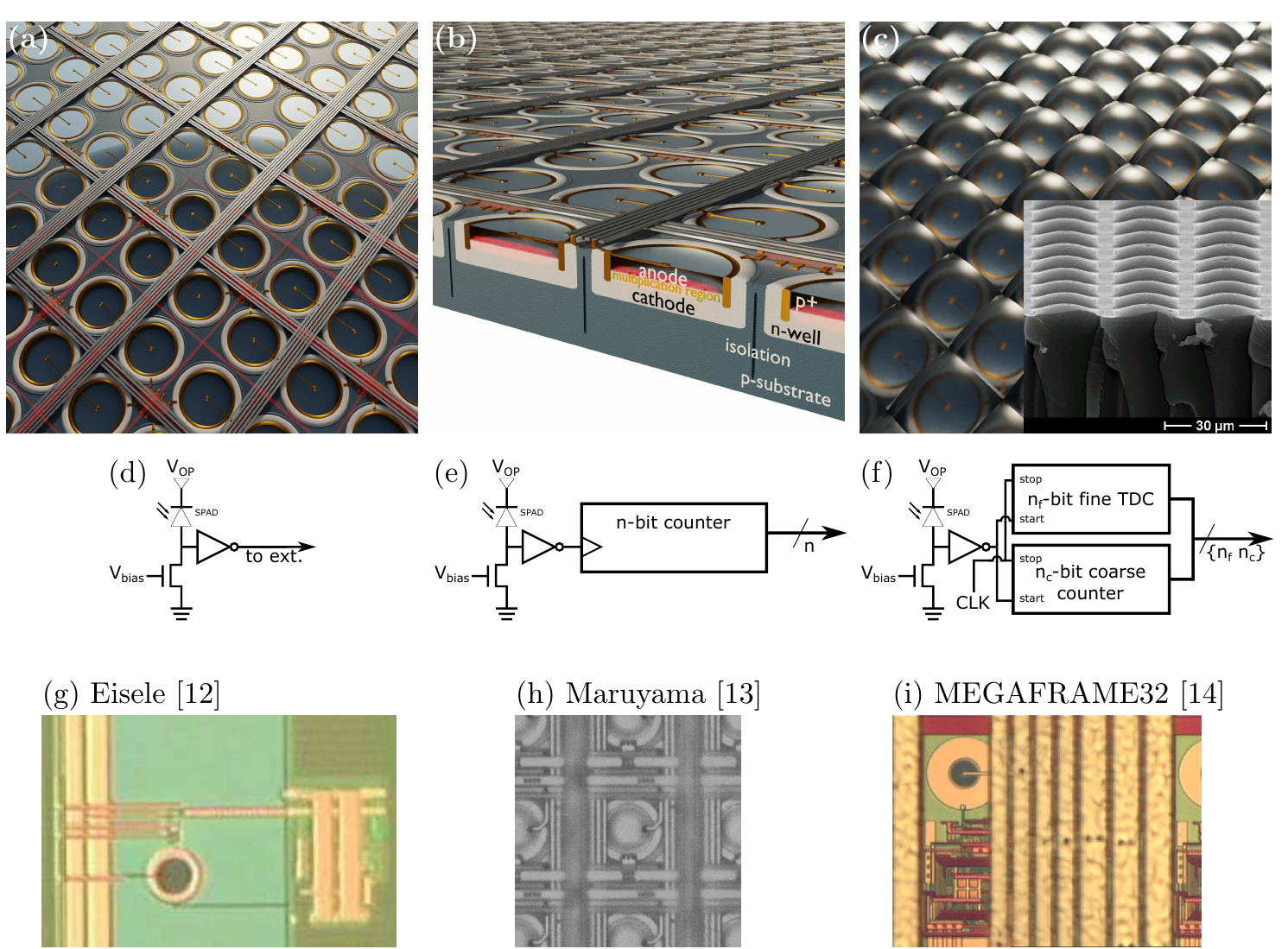}
		\caption{\label{fig:pixel}(a) Artist's impression of a SPAD array (top view) and (b) an example of the corresponding cross section for a substrate isolated SPAD in a conventional CMOS process, depicting some of the key components (diode anode/cathode and corresponding p-n junction, multiplication region in which the avalanche is triggered, substrate and isolation from it) \cite{Charbon2014}. The SPAD fill factor can be enhanced with microlenses (c), with the inset showing a SEM image from \cite{Burri2014_SwissSPAD}. The design of individual pixels ranges from (d) basic structures, only capable of generating digital pulses corresponding to individual photon arrivals on the SPAD, to (e) pixels including counters, which add the individual arrivals over a given time window, possibly gated, or (f) more advanced electronics such as a complete TDC, which make it possible to timestamp individual photon arrival times.
		Corresponding examples of pixel micrographs are displayed in (g--i), reprinted from \cite{Eisele2011, Maruyama2014, Richardson2009}.}
	\end{figure}

Photodectors capable of measuring single photons have been known for decades and have been realized using different technologies, from photo-multiplier tubes (PMTs) to micro-channel plates (MCPs) and electron-multiplying CCDs (EMCCDs). However, the implementation of large, all solid-state single-photon imagers (\autoref{fig:pixel}(a)) calls for a new kind of miniaturized, scalable device featuring a reliable performance parameter set. One example of such a device is represented by the single-photon avalanche diode (SPAD) implemented in industry-standard CMOS technology. The SPAD is basically a photo-diode whose p-n junction (\autoref{fig:pixel}(b)) is reverse biased above its breakdown voltage $V_{bd}$, such that a single photon hitting the active (i.e. photosensitive) device area can create an electron-hole pair and thus trigger an avalanche of secondary carriers. The avalanche build-up time is typically on the order of picoseconds, so that the associated change in voltage can be used to precisely measure the time of the photon arrival \cite{Niclass2005, Zappa2007}. This regime of operation is known as Geiger mode, and hence the devices are also known as Geiger-mode APDs (GmAPDs).

The self-sustaining avalanche in the SPAD needs to be stopped as soon as possible to prevent the destruction of the device itself due to high current. The corresponding quenching takes place by lowering the SPAD bias voltage V$_\text{OP}$ below breakdown, e.g. by using a resistor in series with the SPAD. The voltage across the SPAD then needs to be restored to its initial value above breakdown, before the following photon can trigger another avalanche. During this interval, which is typically on the order of tens of nanoseconds, the SPAD is unable to detect additional photons. This interval is known as dead time. The change in voltage across the SPAD during a detection is easily transformed into a digital signal by using a discriminator, for example a single transistor or an inverter. The resulting output, which does not depend on the wavelength of the impinging photon, is compatible with standard electronics, which makes it easy to integrate a SPAD into larger circuits and arrays of detectors. \autoref{tab:SPAD_key_param} summarizes the most important properties of SPADs and compares them across the SPAD-based imagers reported in \autoref{tab:summary_table}).

\begin{table}
  \centering
  \footnotesize
  \caption{
  \label{tab:SPAD_key_param}Key SPAD pixel parameters and typical values commonly found in the sensors listed in \autoref{tab:summary_table}. }
  \rowcolors{2}{gray!14}{white}
    \begin{tabular}{|l|l>{\hspace{-30pt}}r|c|}
    \hhline{~~~-}
    \rowcolor{gray!30}
    \multicolumn{3}{l|}{\cellcolor{white}}    & Value range \\
    \hline
\cellcolor{gray!30} & Dead time            & [ns]                 & 10--100 \\
\cellcolor{gray!30} & DCR                  & [cps/$\mu$m$^2$]     & 0.6--100 \\
\cellcolor{gray!30} & PDP (peak)           & [\%]                 & 10--50 \\
\cellcolor{gray!30} & Fill factor          & [\%]                 & 1--60 \\
\cellcolor{gray!30} & Timing resolution    & [ps]                 & 30--100 \\
\multirow{-6}[0]{*}{\cellcolor{gray!30}\rotatecol{SPAD pixel}}                     & Afterpulsing probability & [\%]                 & 0.1--10 \\
                     \hline
    \end{tabular}
\end{table}

A number of parameters are used to describe the performance of a single SPAD device \cite{Bronzi2016}. The most important is the photon detection probability (PDP), representing the avalanche probability of the device in response to a photon absorption at a given wavelength. In CMOS SPADs, the PDP has a peak in the visible, which can reach 70\% for single, optimized diodes. Other important parameters are dark count rate (DCR), i.e. the observed avalanche rate in the absence of light, and afterpulsing, which introduces false events that are correlated in time to previous detections. When SPADs are grouped in imagers one has to consider electrical and optical crosstalk, as well as fill factor, which denotes the ratio between the photosensitive area and the total pixel area; the latter does obviously affect the overall imager sensitivity, being multiplied by the PDP to give the overall photon detection efficiency (PDE).

Many of the SPAD characteristics  can be optimized in the design phase, often requiring trade-offs. For example, a larger size of the guard ring, to better contain the high electric field and prevent premature edge breakdown, will positively impact the crosstalk, at the expense of active area and thus fill factor. This can be compensated with larger diodes at the cost of DCR, which increases with diode area. A short dead time allows a higher count rate, and thus a high dynamic range, but increases the afterpulsing probability, which leads to problems when detecting photon correlations. The targeted sensor application should ideally be taken into account during the design phase in order to select the optimal trade-offs, such as sensitivity vs. noise and speed vs. fill factor.

\subsection{From individual SPADs to arrays}

When a suitable SPAD device and pixel circuit have been demonstrated in a given fabrication process, they can be integrated into an array to form a SPAD imager. The simplest array is linear, allowing the designer to place the detection and processing electronics outside the photosensitive area, thus reaching higher fill factors. A 2D array of pixels, on the other hand, needs self-contained circuits, in-pixel or at the periphery, to acquire, store and transmit data. This additional circuitry negatively impacts the fill factor, but eliminates the need for scanning to create a complete image. Some freedom also exists at the level of the spatial granularity; grouping pixels, for example, reduces the overall data throughput, while preserving key information, such as photon timing, and reducing complexity. The same is true for the temporal granularity, allowing, for example, to acquire only a subset of all possible timestamps for specific applications. Finally, the sensor fabrication itself might include post-processing steps, such as the deposition of microlenses to increase the overall sensitivity (\autoref{fig:pixel}(c)).

We will discuss the various architectural choices and the corresponding trade-offs below, moving from the pixel level up to the array design specificities.

\section{Architectures}

\subsection{Pixel architecture}
We divide SPAD pixel circuits into three broad types, depending on the functionality added on top of the basic photon-to-electrical pulse conversion. The first type is represented by a basic structure, which includes only the circuitry necessary for a full detection cycle consisting of avalanche generation, quench and recharge. The output of such a pixel is a train of electrical pulses corresponding to individual photon detections. The second type is a pixel with built-in counter, consisting of a counting circuit and at least one bit of memory; its output is a photon count. The third type of pixel is time-correlated and includes circuitry to discriminate the arrival time of photons; its output can be as simple as a flag for a detection during a given time window, or as complex as a variable number of timestamps reporting distinct photon arrival times. Concept schematics for the three types of pixels are shown in \autoref{fig:pixel}(d--f), while selected implementation examples are displayed in \autoref{fig:pixel}(g--i). The pixel fill factor obtained when assembling an array is inversely proportional to the amount of electronics placed besides the SPAD, which makes it advantageous to use modern fabrication technologies that enable smaller feature sizes.

Pixel design elements common to all types of pixels include active quenching and recharge, masking and gating. Active quenching and recharge can be employed to optimize the detection cycle of a SPAD by reducing the dead time in a well controlled way and thus improving the maximum count rate. In this case, active circuitry can be used to stop the avalanche and recharge it earlier than possible with passive resistive approaches. This limits the amount of charges flowing through the diode, improving its lifetime and reducing afterpulsing. Active techniques for quenching and recharge have been employed in a number of designs \cite{Niclass2005, Niclass2008, Pancheri2009, Stoppa2009, Eisele2011, Maruyama2011, Gersbach2012, Mandai2013, Burri2014_SwissSPAD, Carimatto2015, Krstajic2015, Dutton2016, Homulle2016_FluoCAM}.

Masking is used to selectively disable pixels inside an array. This feature is commonly employed in designs where multiple pixels share circuitry, in order to avoid overloading by particularly noisy pixels, which would otherwise decrease the overall performance. Possible implementations can either switch off the SPAD, thereby preventing avalanches from taking place, or disconnect pixels from the read-out circuit. Switching off the SPAD has the additional benefit of removing possible crosstalk. Examples of pixel architectures with masking are \cite{Maruyama2011, Mandai2013, Carimatto2015, Krstajic2015}.

Gating is another independent element of pixel architectures and consists in enabling the SPAD only for a limited time, down to picosecond ranges. The key advantage of gating is the temporal discrimination that can be achieved through it, even though the overall detection efficiency is reduced. In a setup with repetitive (pulsed) illumination, gating permits to selectively capture photons during a portion of the repetition period, with the added option of shifting the gating window in picosecond steps. This feature can also be used to exclude parts of the photon response that are not of interest. Exemplary gating applications include rapid lifetime determination in FLIM, the accurate reconstruction of a particular optical response in the time domain, the elimination of early/background-related detections, or the reduction of a sample's intrinsic fluorescence in Raman spectroscopy. Gating can also be used for the reduction of DCR by eliminating dark counts occurring outside the time zone of interest. 

Moving beyond basic structures, pixels have been designed to include memory elements for multiple purposes. A single-bit memory can be used to capture a purely binary image during a given time interval (i.e. frame time), for example when it is important to avoid global shutter artifacts; this kind of architecture can be implemented with only a few transistors, thus still allowing to reach reasonable native fill factors (5-10\%), while a fast read-out (10-100~kpfs) is usually employed in order to increase the dynamic range and accumulate multi-bit images off-chip \cite{Maruyama2011, Burri2014_SwissSPAD, Ulku2017}. 
Fabricated in a 130~nm CMOS process, the pixel in \cite{Gyongy2018TED} reaches a notable fill factor of 61\% by using an all NMOS design and analog storage element, as well as deep N-well sharing between pairs of SPAD rows, at the price of a reduction in timing accuracy due to the simplified gating circuit, and a somewhat increased crosstalk between pixels.

Multi-bit and multiple counters allow to differentiate between the number of captured photons. When used together with multiple gates, a simple in-pixel phase detector can be constructed and read-out requirements can be relaxed, while maintaining a good dynamic range. Integrating more memory in each pixel drastically reduces the fill factor and makes it advantageous to move to smaller technology nodes. For example, the 2$\times$ 8-bit counter pixels detailed in \cite{Niclass2009}, implemented in an older 0.35~$\mu$m technology, result in a fill factor of 0.8\%, while the 10-bit counter pixel reported in \cite{Lee2016}, designed in a 0.18~$\mu$m process, results in a fill factor of 14.4\%.

Pixels with integrated arrival time measurement, typically implementing time-to-digital converters (TDCs) or their analog counterparts (time-to-analog converters or TACs), represent the most powerful, but also the most complex pixel architecture. The timing circuitry needs in general to be as compact and low power as possible in order to be integrated in every pixel of an imager, while still offering the required timing resolution. Arrays with in-pixel TDCs do usually not exceed fill factors of a few percent \cite{Gersbach2012, Veerappan2011, Field2013, Field2014}. A ring-oscillator is typically used for timestamping with a resolution in the tens of picoseconds (fine measurement), whereas the timing range is extended with a counter as needed (coarse measurement). Analog techniques, such as in-pixel or column-level TACs or analog memories in the form of capacitors, are making a comeback because they can be implemented in area-efficient ways, at the expense of analog-to-digital converters placed at the periphery of the array or outside of the chip, together with all the difficulties inherent in mixed-signal design, e.g. non-uniformities and mismatch. Notable examples of sensors using analog elements are \cite{Parmesan2015, Perenzoni2016, Perenzoni2016II, Dutton2016}, the last of which presents an array with a fill factor of 26.8\%.

\subsection{Array architecture}\label{sec:Array_architecture}

	\begin{figure}
		\includegraphics[width=\textwidth]{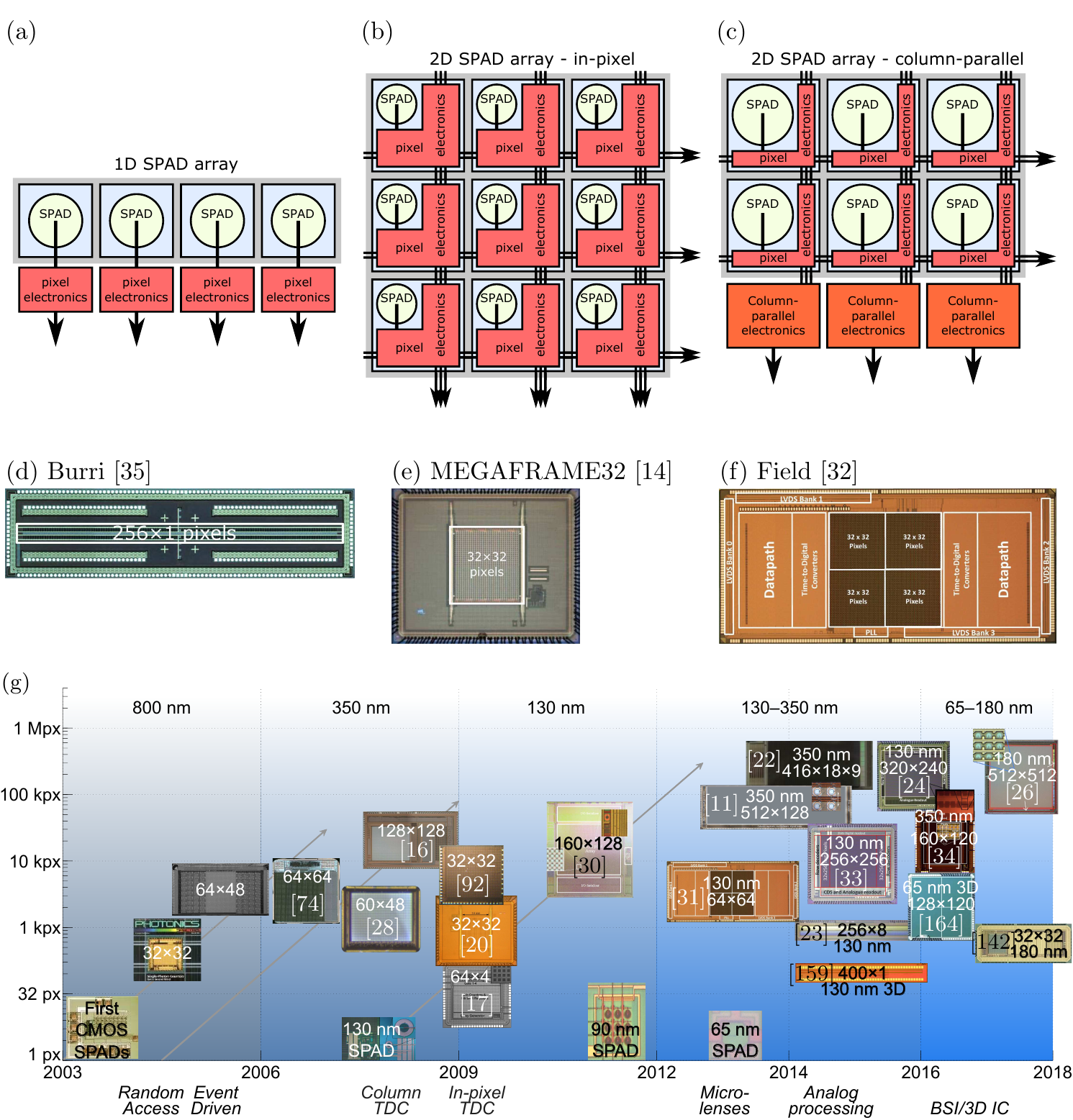}
		\caption{\label{fig:array}Comparison of SPAD array architectures: (a) in linear arrays, the pixel electronics can be placed outside the pixel area, leading to a  fill factor increase; in 2D arrays, fill factors are generally smaller, because (b) electronics is needed inside the pixel itself, or at least (c) at the array periphery, e.g. for column-based TDCs. The related pros and cons are discussed in detail in the text, whereas corresponding examples of array micrographs can be found in (d--f), reprinted from \cite{Burri2017, Richardson2009, Field2014}. Finally, (g) provides an overview of the evolution of SPAD imagers over the last 15~years in terms of total number of pixels (on the vertical axis), the technology node (indicated at the image top), and some salient architectural characteristics, such as Random Access or Event Driven (indicated at the image bottom). Only some representative examples, primarily targeted at biophotonics applications, are shown here (details are reported in \autoref{tab:summary_table}). The diagonal lines indicate developments along a given technology node (800~nm, 350~nm, 130~nm), which usually started by optimizing the SPADs before designing full imagers. The latest years have seen a trend towards higher spatial resolutions and 3D IC solutions.}
	\end{figure}

The simplest form of a SPAD pixel array is a single line. In a linear (1D) array (\autoref{fig:array}(a,\,d)) all pixel electronics is placed outside the sensor area, with only the diode guard ring separating the active area of different pixels. Most linear SPAD arrays allow a truly parallel pixel operation, even if resource sharing is in principle possible in the same way as for 2D arrays. The 1D architecture allows to reach the highest possible fill factors, at the expense of the optical or mechanical scanning solutions that are needed to generate a 2D image, should this be required by the target application.

Two-dimensional SPAD pixel arrays (\autoref{fig:array}(b,\,e)) are capable of acquiring 2D images directly, at the expense of a more complex sensor design for the interconnection between pixels and read-out electronics. In general, all supply, control and data signals are shared across the rows and columns of a 2D pixel array to maximize fill factor. The minimal circuitry needed at the pixel level is a read-out line driver, but usually more is added, such as gating and counters with memory, as discussed in the previous subsection. More complex pixels include timestamping electronics and possibly in-pixel photon information counting or timing pre-processing. Depending on the application requirements, some circuit elements, such as complex time-to-digital converters, can be shared among multiple pixels, either for a larger block of pixels or, more commonly, for (multiple) rows or columns (\autoref{fig:array}(c,\,f)). Non-uniformities and timing skews grow in general with increasing array size, and similarly for the overall generated data volume, calling for specific read-out solutions as discussed below. One possibility is to bin the pixels in groups, e.g. in situations where the spatial resolution can be traded off vs. the signal-to-noise ratio (SNR).

Despite the efforts to maximize fill factors in 2D arrays, the obtained values are usually below those of similarly sized sCMOS cameras (see also \autoref{tab:summary_table}), especially for complex pixel architectures like in-pixel TDCs, due to the larger transistor counts. Microlenses do therefore represent a viable option to reclaim some of the fill factor lost to the electronics. These micro-optical devices are placed in front of the sensitive area, typically on the surface of the detector, and concentrate impinging photons onto the active (i.e. photosensitive) pixel area (\autoref{fig:pixel}(c)). Examples of SPAD-related microlens developments and sensors are presented in \cite{Donati2007, Donati2011, Maruyama2011, Pavia2014II, Intermite2015,  Intermite2015II, Antolovic2016II, Gyongy2018}. The microlenses are typically optimized for specific applications and properties (for example collimation) of the impinging light.

In 2D imagers it is possible that the pixels are no longer strictly operated in parallel, for example when they contain memory elements which are addressed and then reset by the read-out (to gather new photons) on a row-by-row basis (rolling shutter acquisition operation). This can lead to well-known artifacts, such as temporal lag, when recording phenomena with very fast dynamics. Some imagers do therefore foresee true global shutter operation, which provides an image snapshot at a given instant. This can be achieved by activating all pixels together at the start of a frame, to then ``freeze'' data acquisition at the end of the frame and start the read-out operation, with some loss in efficiency (reduced temporal aperture). An alternative which does not call for (many) expensive global signals is represented by an event-driven operation mode, which allows continuous on the fly recording of events as they occur; one way of implementing this is by using a  common bus shared by all pixels (e.g. in a column), together with separate address lines to identify the SPAD that has fired.

Considering all trade-offs, encountered when selecting a pixel and array architecture, there is no single best implementation. The SPAD array's architecture should therefore be chosen based on the target application, sometimes even foregoing classical imaging approaches, or at least profiting from the flexibility that SPAD arrays provide, e.g. by binning pixels and pre-processing data close to the sensor. As an example, we consider time-of-flight positron emission tomography (ToF-PET), where the information of interest is represented by the energy, time-of-arrival and interaction coordinates of gamma rays; the latter are converted by means of scintillating crystals into visible light photons, to be detected by SPAD arrays in the form of SiPMs. In this case, it makes sense to reduce the effective granularity of the recorded data by grouping multiple SPADs together and compensate for noisy detectors using masking. The gamma ray energy is given, in this digital approach, by the total number of SPADs that have fired in a time window of a few hundred nanoseconds, while the time-of-arrival can be estimated on chip and refined on the local control and communication FPGA. An overview of digital approaches to SPAD-based sensors for PET is provided in \cite{Schaart2016}, while individual detector architectures are detailed in \cite{Mandai2013, Walker2013, Braga2014, Carimatto2015}.

\subsection{Read-out architecture}\label{sec:Read-out_architectures}

One of the main concerns when interfacing with a single-photon camera is the resulting high data rate, especially when recording timestamps of individual photons or working at very high frame rates. Eventually the data rate needs to be reduced to a level where it can be transferred to a computer or storage medium. This can be realized, for example, with the same approach as in a streak camera, whereby information captured during a (very) short duration is stored locally (in the pixel) at high speed, and then read out at low speed for processing and storage. Another possibility is represented by \emph{in situ} extraction of higher level information. The corresponding algorithms, like histogram accumulation or multi-bit count integration, can be implemented on the control FPGA, or even on the sensor itself. In the case of fluorescence lifetime imaging, for example, real-time systems have been devised that are capable of calculating the lifetime of molecules at video rate, without the need for recording the full single-photon data stream \cite{Li2010, Li2011, Henderson2014}, including for multi-exponential scenarios \cite{Li2015}. An FPGA system indeed offers some flexibility in terms of possible data processing and high computational bandwidth, which can be used for example to realize firmware-based 32$\times$32 autocorrelator arrays as detailed in \cite{Buchholz2012} (with FCS as target application). The ``reconfigurable pixel'' concept proposed in \cite{Burri2016_LinoSPAD, Burri2017} maximizes flexibility by moving the whole circuitry, which is usually placed beyond the basic SPAD pixel structure, inside the FPGA; this makes it possible to implement different TDC or counter architectures, with the goal of tailoring the system (sensor plus firmware) in an optimal way to the target application's needs.

\subsection{SPAD sensor summary}

\autoref{tab:summary_table} lists a comprehensive summary of the main SPAD-based sensors and imagers which have been targeted at biophotonics applications; they are discussed in the next section, together with the related applications and the corresponding results. A representative subset is shown in \autoref{fig:array}(g), which provides a graphical overview of how these imagers evolved over the last 15~years. \section{Biophotonics applications}\label{sec:applications}

The following sections analyze in detail a host of biophotonics applications that have been explored with SPAD imagers, starting from FLIM, which has been addressed early on, and its different flavors, to end with 
more disruptive scenarios and forward looking sensor concepts.
\section{Fluorescence lifetime imaging}

Fluorescence lifetime imaging (FLIM) offers a non-invasive measurement technique for applications where the fluorescence intensity does not provide sufficient information or discrimination. Common usage of FLIM is found in the study of living tissues and cells at the molecular level, due to the fact that fluorescence lifetime is insensitive to fluorescence intensity and to the corresponding probe concentration, at least to a reasonable extent \cite{Becker2012}; the samples should, however, not be subject to excessive illumination intensities, to avoid phototoxicity and photobleaching. Other advantages are the detection of lifetimes which can be dependent on pH, temperature, oxygen concentration and viscosity levels, thereby enabling the detection of effects that cannot be observed with simple fluorescence intensity measurements.

Slow acquisition speed is the main limitation of standard FLIM setups. While the photo-physics at the molecular level contributes to this, the detection system can impose major speed limitations as well. Time-correlated single-photon counting (TCSPC), which requires the timestamps of individual photons, is often the detection method of choice due to its very high precision, but the underlying hardware and data acquisition are hard to scale to large multichannel arrays; scanning is therefore required when used in an imaging setup. Time-gated sensors have also been employed, including large sensitive areas; they rely on one or more (moving) gates to recover the timing information, and thus the lifetime, at the price of a reduction in the overall sensitivity, as discussed in the Architectures section.

The interested reader is referred to \cite{Lakowicz1992, Marcu2012, Henderson2014, Suhling2015, Hirvonen2016} for background literature on FLIM and related sensors, techniques and applications. We will focus in the following subsections on standard CMOS SPAD implementations for FLIM, and how these have been engineered and employed to address the aforementioned limitations.

\subsection{Point-like FLIM}

Point-like FLIM systems offer increased signal-to-noise ratios by combining the individual outputs of several pixels. An example of such an approach is represented by the FluoCam \cite{Homulle2016_FluoCAM} system, which comprises a 60$\times$48 SPAD array \cite{Niclass2009}, with two 8-bit time-gated counters in each pixel. The gates can be externally programmed to shift in steps of roughly 12~ps, so as to cover a full laser repetition period with high accuracy. The two counters do therefore allow a precise reconstruction of the fluorescence response, even when significant photobleaching distorts the signal, including for sub-nanosecond lifetimes. The integration times are of the order of several minutes, but can be substantially reduced by resorting to more recent designs and/or technology nodes.

\begin{figure}
		\includegraphics[width=\textwidth]{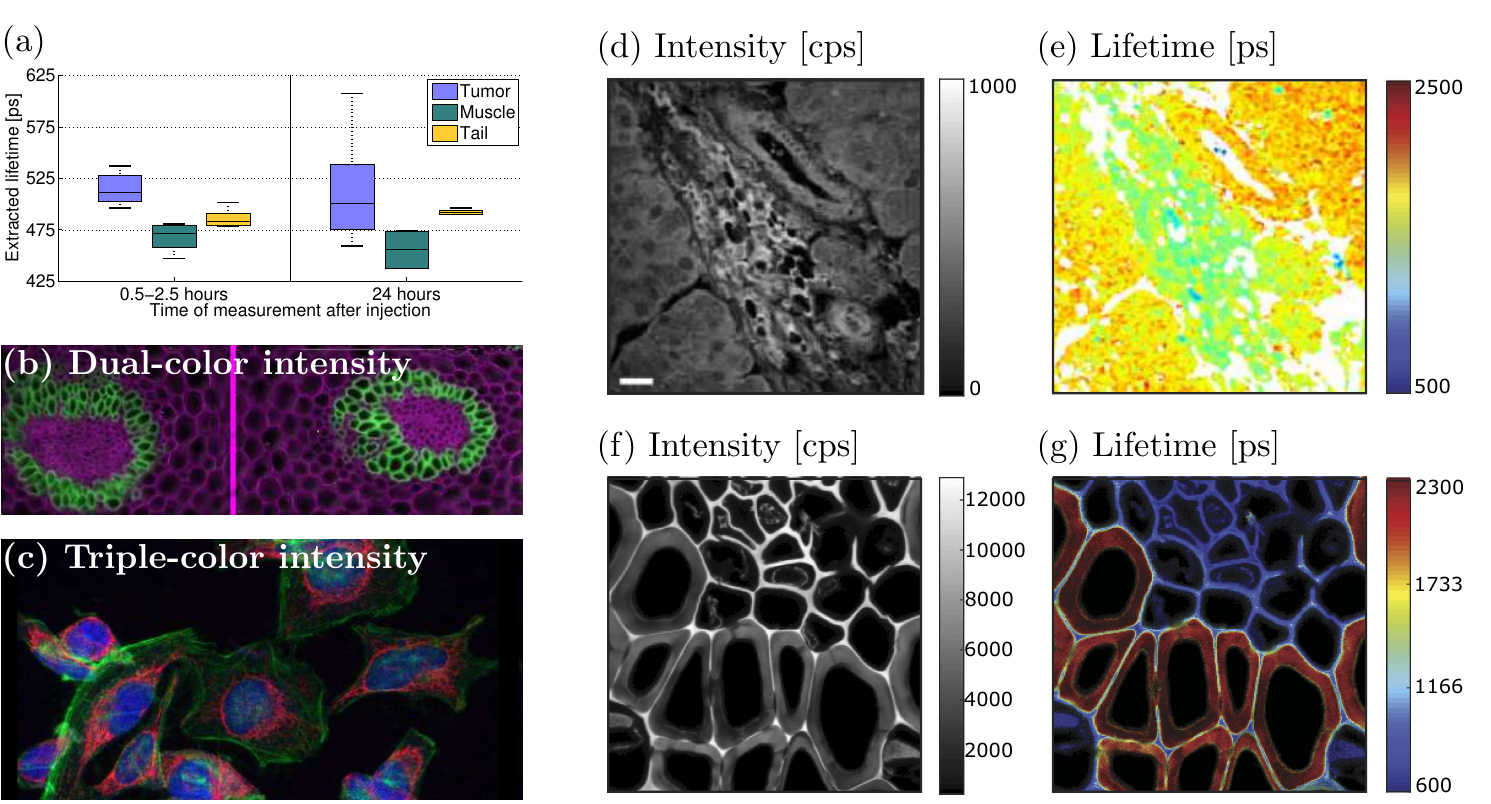}
		\caption{\label{fig:FLIM_image}Examples of fluorescence intensity and/or lifetime results: (a) FluoCam system used in a point-like mode - study of monomeric ICG-\emph{c(RGDfK)} injected in a mouse with a glioblastoma mouse model. A subtle lifetime shift between tumor and non-tumor tissue is observed \cite{Homulle2016_FluoCAM}. (b) Dual-color intensity fluorescence image of a thin slice of a plant root stained with a mixture of Safranin and Fast Green, taken with the SwissSPAD widefield time-domain gated array \cite{Antolovic2016III}. (c) Triple-color intensity fluorescence image of HeLa cells labeled with DAPI, Alexa 488, and Alexa 555, taken with SwissSPAD2 \cite{Ulku2018}. (d,\,e) Label-free FLIM of an unstained liver tissue excised from a tumorogenic murine model \cite{Popleteeva2015}, imaged with a 64$\times$4 SPAD array \cite{Pancheri2009}. (f,\,g) A Convallaria FLIM measurement done with a linear 32$\times$1 SPAD array \cite{Peronio2017}. Images reprinted from \cite{Homulle2016_FluoCAM, Popleteeva2015, Antolovic2016III, Ulku2018, Peronio2017}. 
		}
	\end{figure}

The FluoCam system has been used in several \emph{in vivo} studies to demonstrate the capabilities of such an approach, employing indocyanine green (ICG)-modified derivates, such as ICG-RGD, which target the $\alpha_v\beta_3$ integrin; the final goal was to explore the feasibility of surgical applications with exogenous NIR targeted fluorophores \cite{Stegehuis2015, Homulle2016_FluoCAM}. The system was capable of discriminating between healthy (muscle and tail) and cancerous tissues in a mouse with a glioblastoma mouse model, even though the lifetime difference was only about 50~ps (10\% level in this case) between the lifetimes of the bound and unbound fluorophores as shown in \autoref{fig:FLIM_image}(a).

In TCSPC mode, the detection system's limitations introduce pile-up effects at detection rates $>$0.1 photon/laser cycle, causing underestimation of the lifetime. Pile-up correction using standard SPAD detectors has been discussed in detail by L{\'e}onard \emph{et al.} \cite{Leonard2015}, who show that low count rates are not necessarily needed to avoid pile-up, at the price of an increase in the lifetime estimation variation. In integrated SPAD detectors, multiple subsystems contribute to the pile-up. A detailed analysis of the SPAD, timing and routing dead time influence on lifetime estimation can be found in \cite{Arlt2013}, with experimental results based on a mini-silicon photomultiplier (32$\times$32 SPAD pixels) implemented in a 130~nm CMOS process, and featuring an on-chip lifetime estimator \cite{Tyndall2012I, Tyndall2012II}. The outputs of all pixels are routed towards 16~TDCs that can timestamp up to 8~photons per excitation period. These timestamps are then processed in a center-of-mass module to derive the fluorescence lifetime. This technique has been validated with reference samples with relatively long lifetimes of over 1~ns, demonstrating that reliable lifetimes can be estimated, with a proper architecture, at photon count rates that go well beyond the classical pile-up limit. 

\subsection{Linear SPAD arrays and corresponding FLIM applications}

The pros and cons of linear architectures have already been discussed at the beginning of the Array architecture section. Pancheri \emph{et al.} implemented a 64$\times$4 linear SPAD array (overall size of 1660$\times$104~$\mu$m$^2$ in a 0.35~$\mu$m CMOS technology) targeted for FLIM \cite{Pancheri2009}. The four SPADs in each column were connected to the same read-out channel, creating macro-pixels to reduce the influence of the single SPAD dead time ($\sim$50~ns). This increased the photon throughput of a 15.8$\times$63.2~$\mu$m$^2$ macro-pixel. The chip also featured four time-domain gates that were connected to four separated counters, enabling to construct on-chip histograms of photon-arrival times with four bins, and to compress the data. This sensor was later used \cite{Popleteeva2015} for spectrally resolved FLIM (sFLIM or $\lambda$FLIM), a setup which enables to separate molecules by both the fluorescence emission wavelength and the fluorescence lifetime \cite{Hanley2002}. An example of a corresponding tissue image is shown in \autoref{fig:FLIM_image}(d,\,e). The $\lambda$FLIM system eases the discrimination of different fluorophores and enables to study donor and acceptor molecules simultaneously.

More recently, Krstaji{\'c} \emph{et al.} presented a linear 256$\times$2 SPAD array with pixel pitch of 23.7~$\mu$m and a high fill factor of 43.7\%, implemented in a 130~nm CMOS process \cite{Krstajic2015, Ehrlich2017, Kufcsak2017}. Each pixel was connected to one TDC with a 40~ps LSB. The sensor also featured an on-chip center-of-mass (CMM) calculation for mono-exponential fluorescence lifetimes, enabling to output lifetimes at a 200~Hz line rate with up to 65~kphotons/pixel (limited by the SPAD dead time). Alternatively, the chip can output per-pixel TCSPC histograms with 320~ps bin resolution. Multicolor microspheres and skin autofluorescence lifetimes were measured, with data acquisition of 5~minutes for TCSPC data, to be compared with 2/200~ms when operating in the CMM mode for fluorophores in cuvette/skin autofluorescence, respectively. This chip is currently being used in the \href{http://www.proteus.ac.uk/}{EPSRC UK PROTEUS project}, targeting \emph{in vivo}, \emph{in situ} microendoscopic instrumentation for lung diseases diagnosis. Erdogan \emph{et al.} designed a new generation linear array, in the same 130~nm CMOS process, extending the resolution to 1024$\times$8 \cite{Erdogan2017}. The chip features 512 TDCs and on-chip histogramming that decreases the output data rate and mitigates the I/O and USB bottlenecks.

Linear arrays can also be used as single point detectors, for example by means of optical 1D to 2D transformations, to reduce the effect of single SPAD dead time and increase the throughput in FLIM measurements \cite{Peronio2017}; an example of the corresponding results is shown in \autoref{fig:FLIM_image}(f,\,g). This and other approaches could open the way to high-throughput biotechnological applications, such as high-throughput screening or cell sorting \cite{Leonard2015, Tsikouras2016}, based on nanosecond-lived fluorophores.

\subsection{Widefield time-domain gated FLIM}

Gated SPAD arrays are in principle easier to implement over large areas than TCSPC solutions, and thus provide an appealing path to all-solid-state widefield, time-resolved imaging.

An initial implementation of a time-gated 128$\times$128 SPAD array with 1-bit memory combined an on-chip 600~ps delay line and an off-chip 200~ps delay line for gate shifting \cite{Maruyama2011}. DNA molecules labeled with Cy5 were placed directly on the chip surface and the lifetime was measured. Time-gating enabled excitation elimination without the need for dichroic mirrors.

SwissSPAD \cite{Burri2014_SwissSPAD} is a 512$\times$128 SPAD array with an in-pixel 1-bit memory and time-gating capability. The 1-bit frames are read out at 156~kfps, while time-gating enables independent global exposures as short as 4~ns. The gate position can be shifted in 20~ps steps with respect to a reference signal. This enables to reconstruct exponential lifetimes per pixel. The implemented pixel, which contains 12 transistors, has a 5\% fill factor due to the 0.35~$\mu$m manufacturing process. With the use of microlenses, the effective fill factor is increased to 50--60\% for collimated light, as featured often in microscope output ports \cite{Antolovic2016II}; a representative fluorescence intensity image is shown in \autoref{fig:FLIM_image}(b).

Early characterization of SwissSPAD (without microlenses) for FLIM measurements was performed with reference data sets \cite{Burri2016_thesis}; the sensor was found to be able to properly extract the lifetime of fluorophores in the nanosecond range.  Ulku then designed SwissSPAD's successor, SwissSPAD2 \cite{Ulku2017, Ulku2018}, a 512$\times$512 SPAD array - the largest time-resolved SPAD image sensor to date - with higher PDP and lower DCR, based on a similar architecture. A triple-color fluorescence intensity image is shown in \autoref{fig:FLIM_image}(c). Further research on widefield time-domain gated FLIM with microlense-enabled versions of SwissSPAD architectures is ongoing, including real-time phasor-based measurements \cite{Wargocki2017, Ulku2018}.

Following up on the proof-of-concept work by Pancheri \emph{et al.} \cite{Pancheri2011,Pancheri2012, Pancheri2013}, Perenzoni \emph{et al.} designed a 160$\times$120 SPAD imager with gating, but with multi-bit memory \cite{Perenzoni2016}. The gate can be set as short as 750~ps, with rise and fall times down to 200~ps, and a frequency of 50~MHz. Instead of a 1-bit memory, this chip uses an analog counter, enabling multiple photon accumulations per frame at the cost of introducing ADCs. The pixel pitch is 15~$\mu$m, resulting in a 21\% fill factor in a high-voltage 0.35~$\mu$m CMOS technology. Gyongy \emph{et al.} pursued an n-well shared pixel approach to achieve a high native fill factor of 61\% for a 256$\times$256 SPAD array with 4~ns gates, 600~ps fall times with on-chip delay generator, and a pixel pitch of 16~$\mu$m \cite{Gyongy2018TED}. 

\subsection{Widefield TCSPC FLIM}

A 64$\times$64 40~$\mu$m pitch pixel array designed in 0.35~$\mu$m high-voltage CMOS technology, featuring 64~column-parallel TDCs and a timing resolution of around 350~ps, represented early SPAD TCSPC array work \cite{Schwartz2007, Schwartz2008}. The maximum PDE of 0.1\%, though, remained relatively low.

The MEGAFRAME32 high-performance sensor was smaller (32$\times$32 SPADs) but adopted a radically different architecture, based on 50~ps, 10-bit in-pixel TDCs, working at a maximum rate of 500~kfps \cite{Richardson2009, Gersbach2012}, recording either time-correlated data (one timestamp per pixel), or time-uncorrelated data (6-bit counting). In the former operation mode, up to 0.5~billion timestamps could be generated per second \cite{Krstajic2015II}. The fill factor (1\%) did suffer from the large in-pixel electronics; on the positive side, this did stimulate pioneering microlense work to bring the fill factor back up. MEGAFRAME32 was extensively employed to explore bio-applications and subsequently extended to a 160$\times$128 array (MEGAFRAME128), adding peripheral intelligence (data compression, CMM pre-processing) \cite{Veerappan2011, Veerappan2011ES, Arlt2011}.

Gersbach \emph{et al.} report on early high frame-rate FLIM proof-of-concept investigations \cite{Gersbach2010, Gersbach2012} with MEGAFRAME32. Li \emph{et al.} \cite{Li2010, Li2011, Li2012} illustrate how firmware-based rapid lifetime estimation algorithms, such as CMM (center-of-mass method), make full use of the large number of available timestamps to enable video-rate (50~fps) real-time FLIM operation. An example of the corresponding \emph{in vivo} two photon FLIM data, both intensity and lifetime, is shown in \cite{Li2012} using an FITC-albumin probe, which was injected into a rat bearing a P22 tumor and measured 100~minutes after injection. A clear distinction between the blood vessels and the tumor tissue could be observed in the lifetime image (bi-exponential decay), in contrast to the intensity image. Another widefield FLIM application of the same sensor, coupled to DNA microarrays, is reported in \cite{Giraud2010} employing a TIRF excitation geometry. Distinct lifetime signatures, corresponding to dye-labelled HCV and quantum-dot-labelled HCMV nucleic acid targets, could be distinguished over 320~pixels, with concentrations as low as 10~nM and an exposure time of 26~seconds.

A different architecture was selected by Field \emph{et al.} for their 64$\times$64 array, in a standard 0.13~$\mu$m CMOS process, reported in \cite{Field2013, Field2014}, namely one TDC per pixel (LSB of 62.5~ps) but placed at column level; this led to a pixel pitch of 48~$\mu$m. The sensor was aimed at video rate operation (100~fps), but the corresponding extreme datarate of 42~Gbps entailed a high power consumption (14.5~mW/pixel).

Parmesan \emph{et al.} \cite{Parmesan2015} have chosen to emphasize small pixel pitches (8~$\mu$m), while still maintaining a pixel fill factor of nearly 20\%, by resorting to an architecture based on in-pixel time-to-amplitude (TAC) converters, with a global ramp voltage. This enabled the design of large 256$\times$256 array, which can work either interfaced to external TDCs (optimal timing performance but slower system) or using an on-chip coarse flash ADC (lower temporal resolution).

\subsection{Multibeam FLIM}

Multibeam architectures enable to increase the photon throughput and reduce FLIM acquisition time. Coelho \emph{et al.} and Poland \emph{et al.} used MEGAFRAME32 with Spatial Light Modulators (SLM) for multibeam multiphoton FLIM, increasing the throughput by the number of parallel beamlets \cite{Coelho2013, Poland2013, Poland2014, Poland2015}. Fill factor does not decrease the sensitivity in such setups because beams are concentrated onto the active area of SPADs. FLIM data of live cells (MCF-7 human carcinoma cells) labeled with green fluorescent protein was acquired within 500~ms, at reduced accuracy. This approach was extended in \cite{Poland2016} with a new CMM method, mostly implemented in hardware, capable of pixel level background subtraction and not requiring prior knowledge of the expected lifetime; real-time operation was obtained, at reduced accuracy compared for example with TCSPC mean squared fitting techniques.

Vitali \emph{et al.} also used a multibeam approach with a 32$\times$32 array (square pixels of 100~$\mu$m with a 20~$\mu$m circular SPAD and 8-bit counters) and measured FLIM of eGFP in living HEK293-T cells \cite{Vitali2014}. The SPAD sensor was implemented in a standard CMOS process and included time-gating with minimum gate width of 1.5~ns and delay steps below 100~ps.

\section{F\"orster resonance energy transfer (FRET)}

F\"orster (fluorescence) resonance energy transfer (FRET) uses interactions between two different chromophores (light-sensitive molecules) that non-radiatively transfer energy from a donor to an acceptor molecule. The energy is transferred only when the distance between the two molecules is small enough (nm scale, establishing a long-range dipole-dipole coupling), and when there is sufficient overlap between the emission spectrum of the donor and the excitation spectrum of the acceptor. During this coupling, one can observe a decrease in the donor fluorescence and an increase in the acceptor fluorescence. A typical application is the study of protein-protein interactions and the measurement of distances between molecular groups in protein conformations \cite{Clegg1995}.

FLIM-FRET not only measures the fluorescence intensity change in the donor and acceptor emissions, but also the shortened lifetime of the donor molecule as a result of quenching \cite{Becker2012}. By measuring the ratio between the quenched and non-quenched lifetime, the donor-acceptor interactions can be quantified independently from the molecule concentrations within a diverse sample (in contrast to emission intensity FRET).

Poland \emph{et al.} implemented a MEGAFRAME32-based multifocal FLIM-FRET detector system combined with the optical setup mentioned in the Multibeam FLIM section \cite{Coelho2013, Poland2013, Poland2014, Poland2015}. While scanning protein-protein interactions in live cells with a frame time of 500~ms, they studied changes in FRET interactions between epidermal growth factor receptors (EGFR) and adapter proteins Grb2, as well as a ligand-dependent association of HER2-HER3 receptor tyrosine kinases. Kufcs{\'a}k \emph{et al.} used 5-carboxyfluorescein as a donor and methyl red as an acceptor in FRET for thrombin detection \cite{Kufcsak2017}. Thrombin cleaves the connection between the donor and acceptor, separating them in space and removing the energy transfer.

\section{Fluorescence correlation spectroscopy}

Fluorescence correlation spectroscopy (FCS) measures fluorescence intensity fluctuations in time, with the aim of estimating the concentration and diffusion coefficients of fluorophores, including in live cells. These parameters are extracted from the autocorrelation of the temporal intensity fluctuations. Faster sensors and imagers enable to study smaller and faster molecules \cite{Thompson2002}. In a widefield setup, the correlation between signals from distant volumes measures the direction and velocity of the flow between them.

\subsection{Multibeam FCS}

The multibeam parallelisation principle introduced in the Multibeam FLIM section can also be applied to FCS setups, generating a large number of laser foci using SLMs or diffractive optical elements (DOEs), while taking care to minimize the background signal generated by out-of-focus light. A single SPAD or a group of SPADs are then used to detect fluorescence from each laser focus. Goesch \emph{et al.} used a small, fully integrated 2$\times$2 CMOS SPAD array for the pioneering work on multibeam FCS reported in \cite{Goesch2004}. The multibeam concept was later extended to 8$\times$8 spots to image bright 100~nm diameter fluorescent beads in solutions, using a 32$\times$32 SPAD array \cite{Colyer2011}. The latter was then employed, together with the 32$\times$32 multibeam setup  previously described in \cite{Vitali2014}, to measure FCS of quantum dot diffusion in solution \cite{Guerrieri2009, Guerrieri2010, Vitali2014}. Researchers used a DOE to generate 32$\times$32 spots with a pitch of 100~$\mu$m and a diameter of 12.5~$\mu$m at the image plane (to match the sensor dimensions).

Independently, Kloster-Landsberg \emph{et al.} \cite{Kloster2013} used the 32$\times$32 MEGAFRAME32 sensor to measure multifocal FCS with live cells, employing a frame time of 2~$\mu$s. In this setup, 3$\times$3 laser foci were used for experiments with free eGFP in HeLa cells. A larger multibeam array could not be employed due to high crosstalk between closely packed spots which emerge in this kind of setups.

\subsection{Widefield SPIM-FCS}

FCS coupled to single plane illumination microscopy (SPIM-FCS) enables faster characterization of 3D samples, recording intensity fluctuations over a widefield plane. By illuminating a micrometer-thick light sheet in a $z$-section, out-of-focus light, photobleaching and photodamage can be minimized. First SPIM-FCS data obtained with a SPAD array was presented in \cite{Singh2013}. The work compared the RadHard2 32$\times$32 SPAD array \cite{Carrara2009} with EMCCD and sCMOS cameras. The RadHard2 camera achieved very high read-out speeds, up to 300~kfps, enabling it to extract diffusion coefficients down to 3~$\mu$s, with the best precision compared to EMCCD and sCMOS. The camera was coupled with a real-time 32$\times$32 autocorrelator, implemented on FPGA \cite{Buchholz2012}, which allowed real-time autocorrelation calculation for small molecules in solution, down to 10~$\mu$s diffusion coefficients. However, RadHard2 did not feature microlenses, which obviously affected the overall sensitivity and limited \emph{in vivo} applicability.

\begin{figure}
		\includegraphics[width=\textwidth]{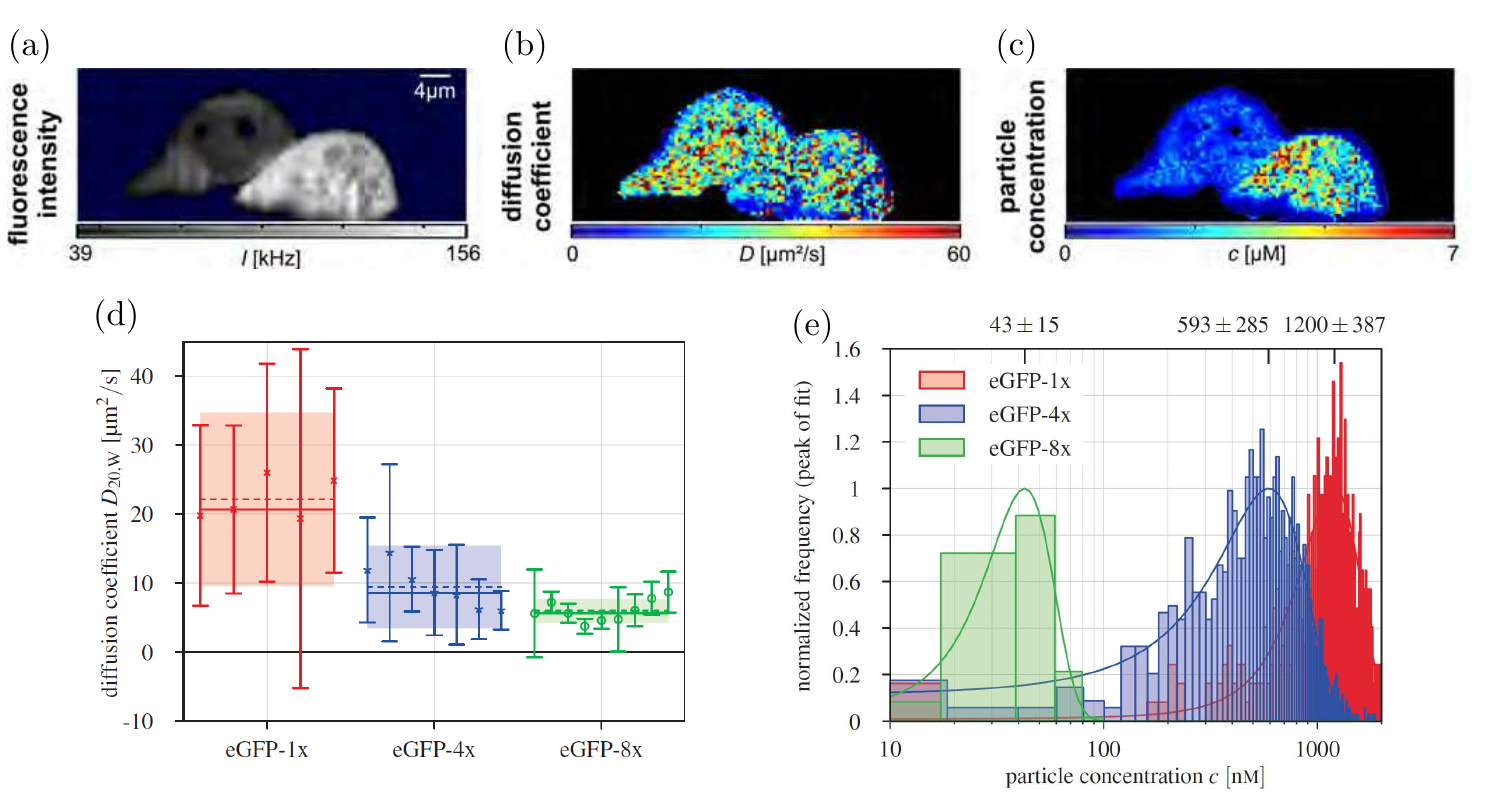}
		\caption {\label{fig:SwissSPAD_FCS}Widefield SPIM-FCS images of monomeric eGFP oligomers in HeLa cells as recorded with a SwissSPAD widefield imager: (a) fluorescence intensity, (b) diffusion coefficient, and (c) dye concentration. (d) Diffusion coefficients for three HeLa cells expressing different oligomers. (e) Particle concentration for the three HeLa cells with different oligomers. Images reprinted from \cite{Buchholz2018}. }
	\end{figure}

Widefield \emph{in vivo} SPIM-FCS with SPAD arrays was firstly demonstrated with a microlensed version of SwissSPAD \cite{Burri2014_SwissSPAD,Antolovic2016II,Burri2016_thesis} by Buchholz \emph{et al.} and Krieger \emph{et al.} \cite{Krieger2014, Krieger2015, Buchholz2018}. FCS results in HeLa cells are shown in \autoref{fig:SwissSPAD_FCS} for three different oligomers of eGFP. The autocorrelation curves of these measurements featured afterpulsing-like increased correlations at short time lags; these artifacts were mitigated by using spatial cross-correlations \cite{Buchholz2016}, which also allowed to determine absolute diffusion coefficients without prior calibration. Although sensitivity needs to be further increased, this work showed that SPAD arrays can measure protein diffusion in live cells with a better SNR than EMCCD cameras and a minimum lag time of $10^{-5}$~s. Correlation algorithms were also extended to GPUs.

\section{Single-molecule techniques}

Single-molecule fluorescence spectroscopy exploits a low concentration regime to excite individual molecules in a very small volume, and collect rare, burst-like fluorescence emission events in correspondence to the transit of individual molecules (whereas in FCS the concentration is such that $\sim$1 or more molecules are present in the excitation volume at any time -- see also \cite{Michalet2014} for a thorough, SPAD-oriented analysis). The number of generated photons is small (a few dozen or less in a large fraction of bursts), resulting in a quite challenging detection. This in turn imposes stringent requirements on the photosensor(s), in terms of sensitivity (including in the red, where standard CMOS is not advantaged due to the low absorption of silicon), fast response time, low noise (read-out or DCR for SPADs), and high count rates, also to be able to separate successive or nearby molecules. Measurement times are usually long, due to the need of accumulating sufficient statistics. 

Several (small) arrays of devices fabricated in custom technologies have been employed over the years with success \cite{Rech2007, Rech2009, Michalet2010, Zappa2010, Michalet2013, Michalet2014, Gulinatti2016}, and parallelization strategies have again been adopted to increase the overall throughput, often employing sophisticated optical setups (multispot excitation and detection). For example, a parallel 8-spot single-molecule FRET (smFRET) analysis system is described in \cite{Ingargiola2012, Ingargiola2013} using 8-pixel SPAD arrays, as well as in \cite{Ingargiola2017}, where the authors tackle realtime kinetic analysis of promoter escape by bacterial RNA polymerase, confirming results obtained by a more indirect route. 

An extension to 48 excitation spots and two 48-pixel SPAD arrays is detailed in \cite{Ingargiola2018}, employing two excitation lasers in order to separate species with one or two active fluorophores. Apart from successfully tackling the multispot setup issues, the resulting smFRET capabilities are shown on a set of doubly labeled double-stranded DNA oligonucleotides with different distances between donor and acceptor dyes along the DNA duplex. The resulting acquisition times are drastically reduced to seconds. This could in turn enable high-throughput screening applications and real-time kinetics studies of enzymatic reactions, potentially propelling single-molecule analysis from a niche technology to a mainstream diagnostics and screening tool \cite{Ingargiola2017}.

\section{Localization-based super-resolution microscopy}

The optical resolution is fundamentally limited by diffraction, whereby Abbe defined the corresponding limit to be $\frac{\lambda}{2n\sin{\theta}}$, where $\lambda$ is the emission wavelength, $n$ the refractive index, and $\theta$ is the half-angle of the cone of light which enters into the objective \cite{Abbe1873}. Several super-resolution techniques have emerged over the years to overcome it, enabling resolution enhancements from initial values of 200~nm down to 10~nm \cite{Hell2009}. One such technique uses sparsely activated (blinking) fluorescent molecules for single-molecule localization \cite{Folling2008}, whereby the location of each molecule is determined by the center of its point-spread function. The final pointillistic image is then reconstructed by merging thousands of frames with hundreds of thousands of localizations, with a resolution limited by the localization precision and labeling density. Due to long frame sequences, tens of seconds to minutes are necessary to reconstruct the final super-resolved image; this time could be shortened by resorting to higher frame rates and/or stronger laser intensities.

	\begin{figure}
		\includegraphics[width=\textwidth]{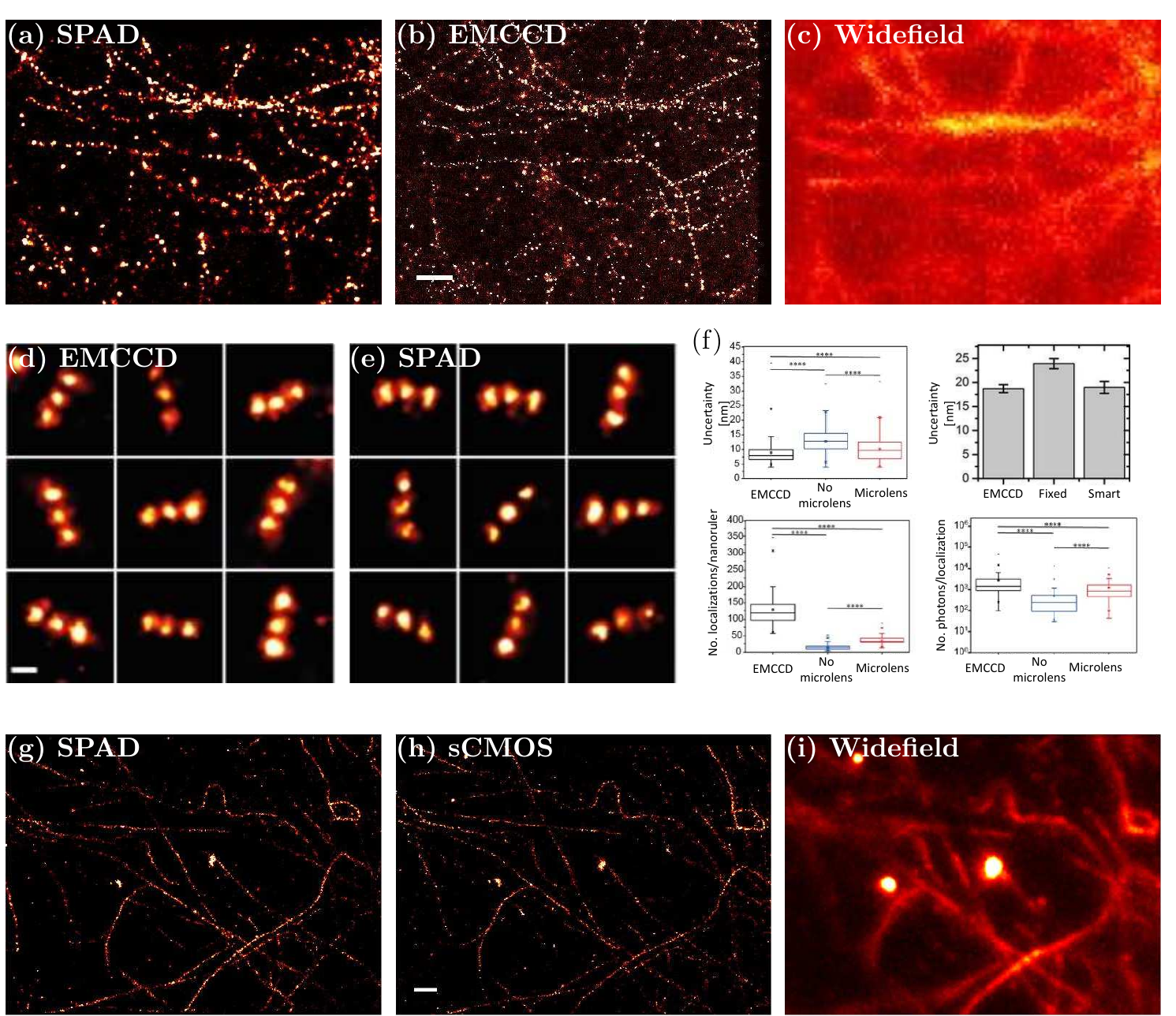}
		\caption{\label{fig:SRM}SPAD super-resolution images: (a) The first super-resolution image captured with SwissSPAD, compared to (b) EMCCD and (c) widefield. Images show microtubuli of an U2OS cell labeled with Alexa Fluor~647, in Vectashield \cite{Antolovic2017}. (d,\,e) Comparison of SPCImager using ``smart'' aggregation and microlenses with an EMCCD. Images show multiple GATTA-PAINT 40G nanoruler localizations \cite{Gyongy2018}. (f) Comparison of differences in localization uncertainty with and without ``smart'' aggregation and the microlenses impact \cite{Gyongy2016,Gyongy2018}. (g) SwissSPAD super-resolution image of microtubuli labeled with Alexa 647 in OxEA buffer, compared to (h) sCMOS and (i) widefield \cite{Antolovic2017}. The white bar shows 1~$\mu$m. Images reprinted from \cite{Antolovic2017, Gyongy2016, Gyongy2018}.
		}
	\end{figure}

Antolovic \emph{et al.}\cite{Antolovic2016} demonstrated the first localization-based super-resolution (SRM) images obtained with SPAD arrays, compared in \autoref{fig:SRM}(a) to EMCCD (\autoref{fig:SRM}(b)) and widefield (\autoref{fig:SRM}(c)), by employing the SwissSPAD imager \cite{Burri2014_SwissSPAD}. Microlenses were deposited on the SPAD array to improve the fill factor from the native 5\% to 60\% \cite{Antolovic2016II}, which was a key enabler for sensitivity-critical applications. The image resolution was analyzed with the Fourier ring correlation method \cite{Nieuwenhuizen2013}, yielding around 100~nm resolution. The estimated localization uncertainty \cite{Thompson2002Loc} was 30~nm, with 200~photons per localization, compared to 15~nm obtained with an EMCCD camera, using 1800~photons per localization. In other terms, although almost 10$\times$ more photons were acquired with the EMCCD camera, the localization results were only 2$\times$ better; the reason is that the SNR, which should indeed increase by $\sqrt{10}$, is reduced by $\sqrt{2}$ due to the multiplication noise in the EMCCD camera itself, which arises from the electron multiplication process. Later results with an improved buffer \cite{Nahidiazar2016}, shown in \autoref{fig:SRM}(g--h), yielded 800~photons per localization with a localization uncertainty of 10~nm for an sCMOS camera, while SwissSPAD collected 100 photons under the same conditions, leading to an uncertainty of 20~nm \cite{Antolovic2017}. 

When compared to standard EMCCD and sCMOS cameras, SPAD imagers eliminate read-out noise by utilizing a direct photon-to-digital conversion -- the digital nature of the SPAD imagers enables fast and noiseless read-out. These properties of SPAD imagers were used in three ways. First, Gyongy \emph{et al.} \cite{Gyongy2016} used temporal oversampling and ``smart'' aggregation to determine the start and end of the fluorophore blinking in a more accurate way, so as to minimize the background noise. Second, Antolovic \emph{et al.} \cite{Antolovic2017} investigated the performance of the localization algorithm at different frame rates. Finally, SPAD imagers were used to perform a widefield analysis of fluorophore blinking in the $\mu$s range \cite{Antolovic2017}.

Concerning the first point, Gyongy \emph{et al.} used the 320$\times$240~SPAD SPCImager \cite{Dutton2016} for localization-based super-resolution imaging \cite{Gyongy2016,Gyongy2018}. By employing microlenses, they improved the effective fill factor from 27\% to 50\% and used this sensitivity enhancement to demonstrate 40~nm resolution. A GATTA-PAINT 40G nanoruler was used as a reference, with the corresponding results shown in \autoref{fig:SRM}(d--f). ``Smart'' aggregation decreased the localization uncertainty by 20\%; simulations show a potential improvement by 50\%. Although the SPAD results have a localization uncertainty comparable to EMCCDs, a 3$\times$ lower sensitivity in the green entails a lower number of localizations.

Concerning the second point, Antolovic quantified the optimum frame rate given the exponential distribution of fluorophore blinking \cite{Antolovic2017}, if oversampling and smart aggregation cannot be used. Oversampling leads to higher localization uncertainty, while undersampling causes a drop in the number of localizations. 

Finally, SwissSPAD's high frame rate of 156~kfps was used to explore additional high-frequency $\mu$s blinking, which reduces the total number of collected photons and thus the final resolution. As an example, the percentage of high-frequency blinking Alexa 647 molecules decreased from \textgreater68\% to \textless30\% when switching from a MEA to an OxEA buffer \cite{Antolovic2017}. This indicates that high speed SPAD imagers could be beneficial for fluorophore/buffer design and optimization.

As the previous discussion has indicated, the fundamental differences between EMCCD, sCMOS and SPAD imagers do also need to take into account, in addition to the overall sensitivity, the different noise contributions and achievable frame rates. A theoretical comparison between these imagers, when applied to localization based super-resolution microscopy, has been conducted by Krishnaswami \cite{Krishnaswami2014}. EMCCDs feature 90\% quantum efficiency, but also multiplication noise that effectively reduces this value by half \cite{Huang2013}; their noise performance is very uniform, due to the serial read-out, but the achievable frame rate is low. sCMOS parallelizes the read-out and increases the frame rate, at the expense of mismatches in the analog electronics. SPAD imagers eliminate read-out noise with fully parallel digitization and can thus achieve very high frame rates for binary (1-bit) frames. If multiple 1-bit frames are added to achieve an 8 to 12 bit depth, which might be necessary or not depending on the application, their frame rates become comparable to the sCMOS ones. On chip counters will eliminate this bottleneck, enabling high frame rates with higher bit depth. In addition, the high speed and picosecond temporal resolution of TDCs connected to SPADs can open new avenues such as video-rate localization super-resolution and multicolor imaging. The SPAD digitization also limits the noise contributions solely to DCR (and afterpulsing), and although DCR levels are acceptable for super-resolution applications, the percentage of ``hot'' pixels needs to be further reduced.

\section{Raman spectroscopy}

Raman spectroscopy is capable of providing data on the chemical composition and molecular structure of a compound in a non-destructive and label-free way \cite{Colthup1975, Krafft2012, Shipp2017}, with applications to both \emph{in vitro} and \emph{in vivo} tissue diagnostics, minimal preparation, and without damaging the samples. This ``fingerprinting'' technique relies on inelastic light scattering with vibrating molecules, whereby Stokes-scattered Raman photons are red-shifted. It has seen a surge of biophotonics applications in the last couple of decades \cite{Krafft2012}, although it typically suffers from a weak scattering cross section, leading to long overall acquisition times (unless used for example in combination with other sampling techniques that reduce the area to be interrogated, employing line scanning, or multi-focal/widefield setups). As such, the Raman signal is often overshadowed by the sample (auto)fluorescence itself; when working with biological samples, it might therefore be of interest to move towards the NIR range, where autofluorescence is weaker, although the resulting SNR needs to be weighted by the corresponding reduction in the scattered intensity, which decreases with the 4$^{th}$ power of the excitation wavelength. The use of coherent techniques such as nonlinear anti-Stokes Raman Scattering (CARS), resulting in blue-shifted radiation, can enhance the scattered signal and circumvent these limitations, enabling for example rapid chemical imaging, at the price of appropriate tunable, femtosecond laser sources \cite{Krafft2012}. Penetration in deep tissues, which is of particular relevance for \emph{in vivo} medical diagnosis or when looking at live cells growing in 3D cultures, can be enhanced with recently developed methods such as spatially offset Raman spectroscopy (SORS) and Transmission
Raman Spectroscopy (TRS); the interested reader can consult \cite{Matousek2016} for a recent review.

A downside of moving to the NIR to reduce the (auto)fluorescence background is that the sensitivity of standard CCD/CMOS imagers decreases, and this is obviously all the more true for the SPAD imagers in standard CMOS onto which this review is primarily focused. However, SPADs allow the design of compact, all solid-state detectors for Raman spectroscopes operating in time-resolved mode, whether via very short gates, ideally in the 10-100~ps range given the nature of the Raman signal (very fast emission compared to the fluorescence background, which is typically in the ns range) \cite{Nissinen2011, Kostamovaara2013}, or based on timestamping, e.g. with TDCs; this mode of operation also allows to reduce the DCR and enhance the overall SNR. Several linear SPAD-based systems have indeed been developed in the recent past, usually comprising one or a few lines, targeting initially applications such as mineralogy, and subsequently biophotonics as well.

One of the largest reported SPAD arrays is a 1024$\times$8 system \cite{Blacksberg2011, Maruyama2013, Maruyama2014} which was applied with success to mineral samples, where the background fluorescence is longer than for typical biological samples of interest to us. It featured 1-bit counters and a global gate as short as 750~ps (with a corresponding standard deviation of 120~ps), which could be shifted in 250~ps steps. The use of a standard 0.35~$\mu$m CMOS process led to a maximum PDE of 9.3\% with a median DCR of 5.7~kcps at 3~V of excess bias. 

A different approach was chosen in \cite{Nissinen2011, Nissinen2013, Nissinen2015}, where a 2$\times$(4)$\times$128 SPAD array, again in a standard 0.35~$\mu$m CMOS process, featured 4 sub-ns time gates during which the on-chip electronics counts the number of detected photons. This architecture also allowed to determine, in addition to the Raman signal, the level of the residual fluorescence and DCR. The fill factor of each SPAD was 23\%, the minimum time gate width 80~ps, and its variation along the spectral axis (timing skew) $\pm$17.5~ps for a nominal width of 100~ps; the effect of nonhomogeneities of this order on the samples of interest for biophotonics applications (featuring ns- and sub-ns scale fluorescence backgrounds), although small, were discussed as well. The overall IRF (instrument response function) was reported as 250~ps, and Raman spectra of several drugs of interest were acquired with a ps pulsed laser at 532~nm \cite{Rojalin2016}, enabling the authors to reveal previously unseen Raman spectral features. The same underlying standard CMOS technology was used to design a gated 4$\times$128 SPAD array featuring an additional 512-channel 3-bit flash TDC \cite{Nissinen2015II}, which allowed high timing resolution measurements (78~ps, 10~ps standard deviation in the first four bins) at the beginning of the 3.5~ns range which it covered. Several SPADs were again employed at a given spectral point, to reduce the impact of noisy pixels.

Further nonhomogeneity studies have been carried out by the same group, including by means of a 256$\times$16 SPAD test array with two on-chip TDCs, leading to a detailed investigation of the effect of DCR and PDE as well as gate length variations and temporal skews. The latter can indeed play an increasing role, with growing design complexity, on the derivation of the Raman spectrum when targeting 100~ps accuracy levels \cite{Holma2017, Nissinen2017}. In general, the spectra have been found to be subject to a larger deterioration, as could be expected, when the fluorescence lifetimes and levels get shorter and higher, respectively. An efficient post-processing method relies on the calibration with a known smooth (over the wavelength) fluorescence background; this approach is discussed in \cite{Nissinen2018}, using a high-precision TDC ($<$ 50~ps) whose range exceeds the expected timing skew, and sampling as well a part of the fluorescence spectrum.

The linear 256$\times$2 array by Krstaji{\'c} \emph{et al.} \cite{Krstajic2015}, which was already mentioned in the FLIM section, has also been used for surface-enhanced Raman spectroscopy (SERS), again within the \href{http://www.proteus.ac.uk/}{PROTEUS project}. Significant hardware and software improvements of the same sensor are reported in \cite{Kufcsak2017}, whereas removal of fluorescent background signals (in addition to DCR) is illustrated in \cite{Ehrlich2017}.

\section{Optical tomography}

Near-infrared optical tomography (NIROT) studies absorption and scattering of light in turbid media, e.g. biological tissue. Light that has propagated in the medium is detected at its surface. The measurement process is performed at multiple wavelengths to exploit the knowledge of the absorption and scattering properties of the constituent absorbers. This prior knowledge can then be used to make a 3D reconstruction of the distribution of those absorbers within the medium; this enables, for example, the non-invasive determination of oxygenation values, usually by working in the 650-850~nm window (enhanced tissue penetration).

The corresponding image reconstruction inverse problem is, however, ill-posed by nature, leading to a low spatial resolution (1~cm level) with existing PMT-based NIROT system; these have seen little use in clinical settings, and are difficult to upscale. The spatial resolution can be improved by increasing the measured amount of information, i.e. ideally the amount of detectors, as well as their temporal information in the case of time-domain systems. This has led to interest in solid-state detectors and in particular in SPADs \cite{Alayed2017}, either as single devices, or in the form of SiPMs, potentially enabling contactless setups and optimized reconstruction algorithms. 

While SiPMs can indeed be combined to form a wide area detector \cite{Ferocino2018}, separate TCSPC electronics is still required, and cost might also play an issue. Another route has therefore explored fully integrated CMOS SPAD arrays featuring high PDE and timing accuracy, as well as thousands of detection channels. An overview of the topic is provided for example in \cite{Pavia2014, Pavia2014II, Pavia2015_thesis}, which also include application examples based on the LASP 128$\times$128 TCSPC array \cite{Niclass2008}. The latter employed 32 10-bit 98~ps column-based TDCs in a pulsed (time-domain) setup, recording time-of-flight information.

\begin{figure}
		\includegraphics[width=\textwidth]{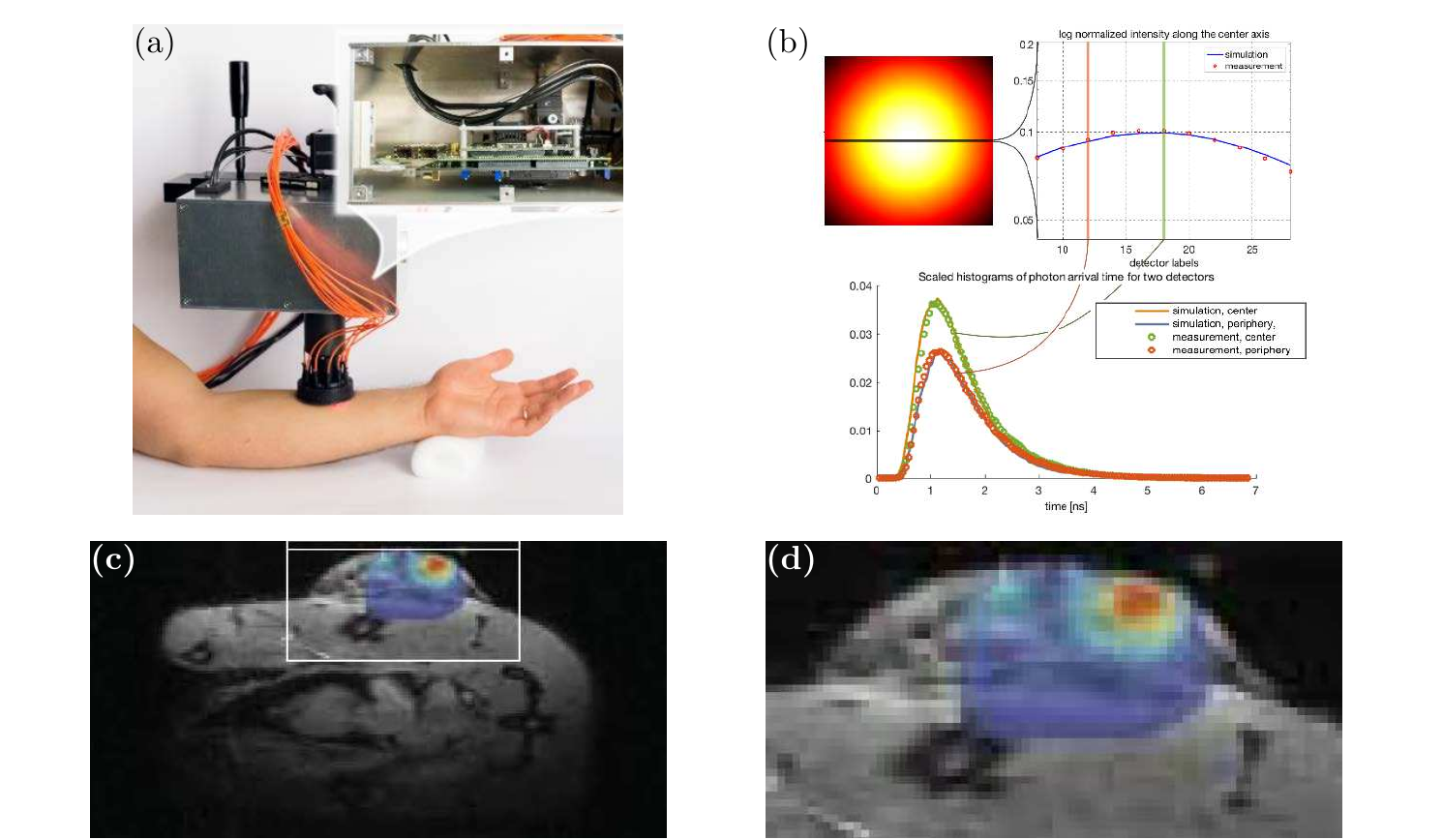}
		\caption{\label{fig:RadHard2_NIROT}(a,\,b) NIROT camera system prototype and measurements vs. simulation results for a phantom \cite{Kalyanov2018}. (c,\,d) Fluorescence molecular tomography (FMT) image as overlap of the optical image obtained with the RadHard2 32$\times$32 photon-counting sensor with the corresponding MRI image \cite{Stuker2011}. C51 cells (a colon cancer-derived cell line) have been implanted in the flank of a mouse. A clear spread in the protease activity, indicated by the significant higher fluorescence intensity in some parts of the tumor, is shown. (c) Complete MR+FMT image, (d) zoom on cancer region. Images reprinted from \cite{Kalyanov2018, Stuker2011}.}
	\end{figure}

An enhanced CMOS SPAD sensor optimized for NIROT is described in \cite{Lindner2017, Lindner2018}, targeted at addressing the slow acquisition time bottleneck of previous implementations, which can lead to motion artefacts and decreased patient comfort. This 32$\times$32 SPAD array, called Piccolo, employs 128 TDCs, based on the concept of dynamic TDC reallocation to reduce the timing-related die area whilst at the same time reducing the probability of photon pile-up. It is implemented in a 180~nm process employing a large spectral range SPAD and cascoded passive quenching circuit to boost the pixel's PDE, reaching a PDP of greater than 10\% at 800~nm with a native fill factor of 28\%, at a pixel pitch of 28.5~$\mu$m. Initial results indicate a minimum image acquisition time of 3.8~s per source and wavelength pair. The sensor has been integrated into a system combining a supercontinuum laser and an acousto-optical filter for multi-wavelength excitation, as well as a fiber switch to obtain up to 24 source positions \cite{Kalyanov2018} (\autoref{fig:RadHard2_NIROT}(a)); phantom measurements are in good agremeent with simulations (\autoref{fig:RadHard2_NIROT}(b)). Future implementations could make use of a SPAD gating feature \cite{DallaMora2015}, in order to reduce the impact of the early backscattered photons in certain illumination geometries.

Fluorescence molecular tomography (FMT) relies on (exogenous) fluorescent molecules to enable the reconstruction of their concentration distribution in 3D, e.g. applied to small animals for pre-clinical cancer research, and possibly over time to track dynamic phenomena. A hybrid MRI-FMT imaging system has been demonstrated in \cite{Stuker2010, Stuker2011}, with the imaging component based on the RadHard2\cite{Carrara2009} 32$\times$32 photon-counting SPAD array. The results are shown in \autoref{fig:RadHard2_NIROT}, whereby an overlap of the MRI and FMT images of a mouse tumor is displayed in (c) and enlarged in (d). The corresponding findings were later confirmed by the histological analysis on the tumor.

Finally, it is worthwhile to mention the possibility of using the exquisite timing capabilities and spatial resolution of SPAD arrays to detect the ballistic and snake photons which have travelled through tissue, such as demonstrated in \cite{Tanner2017}. The MEGAFRAME32 sensor was employed to locate the distal-end of a fibre-optic device deep in tissue, with cm resolution and in clinically relevant settings and chest and lung models, as well as through the entire thickness of a human torso.

\section{Other biophotonics applications and sensor concepts}

Amongst the several other biophotonics applications which have been the subject of investigation with SPAD arrays, the nuclear medicine domain, and PET in particular, play a prominent role. Its peculiar architectural implications have already been analysed at the end of the Array architecture section, and are particularly interesting because they basically span the full pixel granularity range, from small (mm sized) silicon photomultipliers (SiPMs) where all SPADs are connected together to provide a common output, to imagers with individually addressable SPAD pixels. The former, which are already employed in analog form in top-end PET systems, are being revisited, e.g. monolithically with the addition of on-chip TDCs, while keeping backward compatibility \cite{Muntean2017}, or by coupling an off-the-shelf analog SiPM array to an FPGA-based board to enable advanced timing functionality with relatively simple hardware modifications \cite{Venialgo2017}. The latter are being extended in 3D, as will be described in the corresponding section below.

On the more future oriented side, the SPAD's spatial and time-resolved capabilities are being investigated to enable quantum-based superresolution microscopy. In principle, using a quantum correlated N-photon state combined with an optical centroid measurement (OCM) allows indeed to reach a resolution enhancement of 1/N. Apart from the significant challenges on the source side, such an approach calls for high detection efficiency, timing resolution and frame rates on the detection side, while minimising crosstalk and DCR. An early implementation of a monolithic 4$\times$4 G$^{(2)}$ SPAD imager was reported in \cite{Boiko2009, Boiko2009II}, aimed at resolving second-order intensity correlations. More recently, the SUPERTWIN project started looking at all solid-state technologies for the generation and recording of entangled photons, targeting a resolution of 40~nm. The much larger and advanced SPADnet1 SPAD array, originally designed for PET applications, was employed in a first proof-of-principle experiment to detect spatially entangled photon pairs, generated by spontaneous parametric down-conversion (SPDC) in a non-linear crystal pumped with an intense laser beam \cite{Unternahrer2016, Gasparini2017}. The collected experience allowed to proceed with the design of an \emph{ad hoc} detector, incorporating on-chip features to increase the duty cycle (e.g. avoid reading empty frames). A 32$\times$32 SPAD array was manufactured in a 150~nm CMOS process, allowing 50~ns long observation windows at up to 800~kHz to measure 1st (G$^{(1)}$) and 2nd (G$^{(2)}$) order correlation functions, and experimentally characterized again with a source of entangled photon pairs \cite{Gasparini2018}.

A sensor concept which is alternative to the monolithic approaches described so far, built around the use of ``reconfigurable pixels'', has already been hinted at at the end of the Read-out architecture section. The underlying idea is to be able to reconfigure the main data processing features as a function of the specific application needs, as is for example the case with the LinoSPAD 256$\times$1 linear array. This system is designed in such a way that the sensor layer hosting the actual SPADs (with fill factor of 43\%) is decoupled from the data processing features, which are embedded in a companion FPGA. The latter contains core processing blocks such as a 64-element parallel TDC array \cite{Burri2016_LinoSPAD, Burri2017} for time-correlated applications, whereas others are reconfigurable and can in principle be combined in a modular fashion. Another advantage of this approach resides in the possible combination of a sensing-optimized layer with a more advanced processing tier (e.g. 40~nm or 28~nm), as well as the possibility of exchanging one or the other as technology evolves or new firmware is designed. 

	\begin{figure}
		\includegraphics[width=\textwidth]{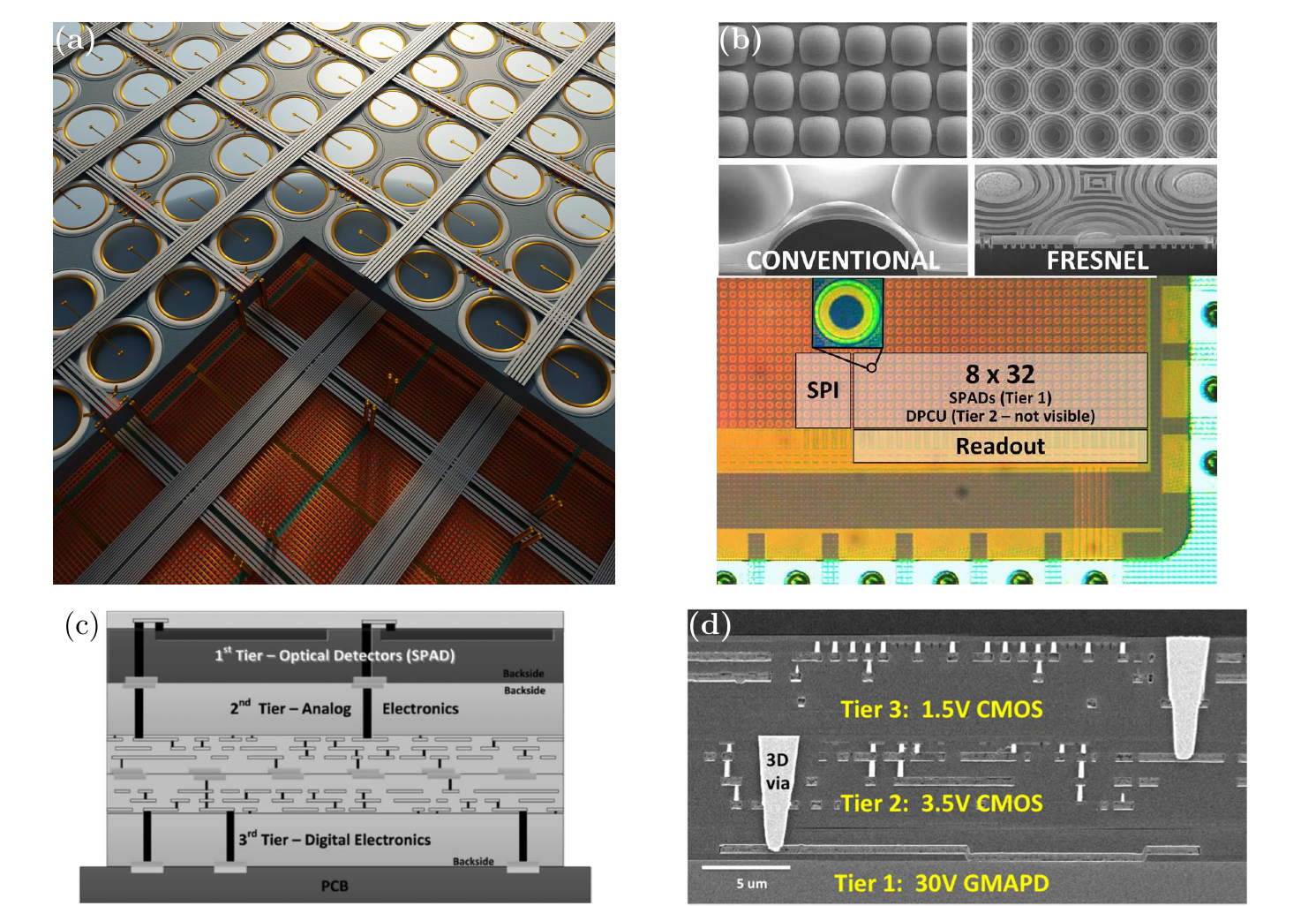}
		\caption{\label{fig:SPAD_imagers}Recent SPAD concepts for imagers revolve around 3D integration, possibly combined with microlenses to further maximize the fill factor. (a) A 3D integration concept image, (b) a two-tier implementation with additional microlenses \cite{Ximenes2018}, and (c,\,d) cross sections of different imagers using three tiers \cite{Aull2016, Berube2015}. Frontside illumination is used in (c), backside illumination in (b) and (d). (b--d) Images reprinted from \cite{Ximenes2018, Aull2016, Berube2015}.}
	\end{figure}

The combination of a top sensor layer with a bottom (likely all-digital CMOS) control and processing layer, each optimized for the respective function, can also be achieved with 3D-stacking techniques (see \autoref{fig:SPAD_imagers}(a) for a concept image), which are progressively getting accessible to a larger user community and which profit from the developments in consumer markets (e.g. cameras for mobile phone applications), where significant resources are available. Such an approach could potentially enable to reach high PDE, low DCR and reduced jitter and afterpulsing, while at the same time adding advanced functionality and low power consumption due to the use of smaller technology nodes in the bottom tier. In addition, the use of compound semiconductors for the top layer, such as InGaAs, could open up additional imaging windows in the NIR and even mid-wave infrared, leading to enhanced tissue penetration (see for example \autoref{fig:SPAD_imagers}(d) and \cite{Aull2015, Aull2016} for a summary of extensive work in this direction by MIT's Lincoln Laboratory). Another route to enhanced NIR sensitivity when staying with silicon-based platforms is backside illumination (BSI), similarly to what already implemented in most high-end CMOS consumer imagers, whereby the substrate of the top wafer needs to be thinned down to only a few micrometers; this can be combined with thicker active volumes to counteract the reduced absorption of silicon in the NIR. In general, the price to pay for going 3D is higher design complexity, challenging 3D (wafer level) bonding techniques, and corresponding development costs.

Early work towards CMOS 3D IC SPAD-based imagers includes a proof-of-concept 400$\times$1 linear array \cite{Pavia2015, Pavia2015_thesis}, specifically designed with NIROT applications in mind, which combined 400~SPADs on the top tier with 100~TDCs (50~ps LSB) on the bottom layer, in a 130~nm process. In parallel, the use of silicon-on-insulator (SOI) was explored by \cite{Durini2012, Zou2014}, with the aim of fabricating an array of 32$\times$32 custom (BackSPAD) photodetectors in a 3~$\mu$m thick SOI film of an SOI wafer (0.35~$\mu$m feature size). The latter is then flipped and wafer bonded to a second 0.35~$\mu$m CMOS wafer with the ancillary electronics. A large body of work is available from Sherbrooke on 3D specifically dedicated to PET, with the aim of achieving a possible ``ultimate'' one-to-one coupling between each pixel and its corresponding quenching and timing electronics \cite{Nolet2016, Nolet2017}. An important part of the effort was dedicated, not surprisingly, to the 3D integration process itself, with proofs of principle along the way such as \cite{Berube2015} detailing the design of the top tier 0.8~$\mu$m high voltage CMOS SPADs in a frontside illumination approach (FSI, see \autoref{fig:SPAD_imagers}(c)), and the corresponding challenges in terms of connections (use of through silicon vias).

More recently, several papers report on work exploiting very advanced commercially available 3D BSI technologies. They were in part driven by the need for optimized, small pitch SPAD arrays with enhanced red and NIR sensitivity for consumer applications such as LIDAR, and have led to SPAD implementations that were unthinkable in such technology nodes just a few years ago, also allowing potential extension to megapixel arrays. In \cite{Abbas2016} for example, the pixel pitch was reduced to 7.83~$\mu$m leading to a 128$\times$120 time-gated prototype combining an imaging specific 65~nm top tier with a 40~nm bottom tier employing 1-to-1 hybrid bond connections. A different implementation in the same 3D IC CMOS process is reported in \cite{Lindner2017EDL}, achieving higher PDE thanks to cascoded passive quenching and active recharge, while still being compatible with the transistor operating voltage regimes of such highly scaled technologies. The example in \autoref{fig:SPAD_imagers}(b) employs a 45~nm sensor tier and 65~nm low-power processing tier, adding microlenses for fill factor enhancement. Very low afterpulsing was obtained with a very short 8~ns dead time. Finally, the top tier was further scaled down to 45~nm in \cite{Lee2017}, while further ameliorating the peak PDP and reducing the jitter and DCR. The latter stays in general 2-3 orders of magnitude higher than the best FSI technologies, whereas the PDP, which is typically very low in the blue region for BSI implementations, has improved to above 30\% in the red, at fill factors well in excess of 50\%.

\section{Summary}
\begin{figure}
		\includegraphics[width=\textwidth]{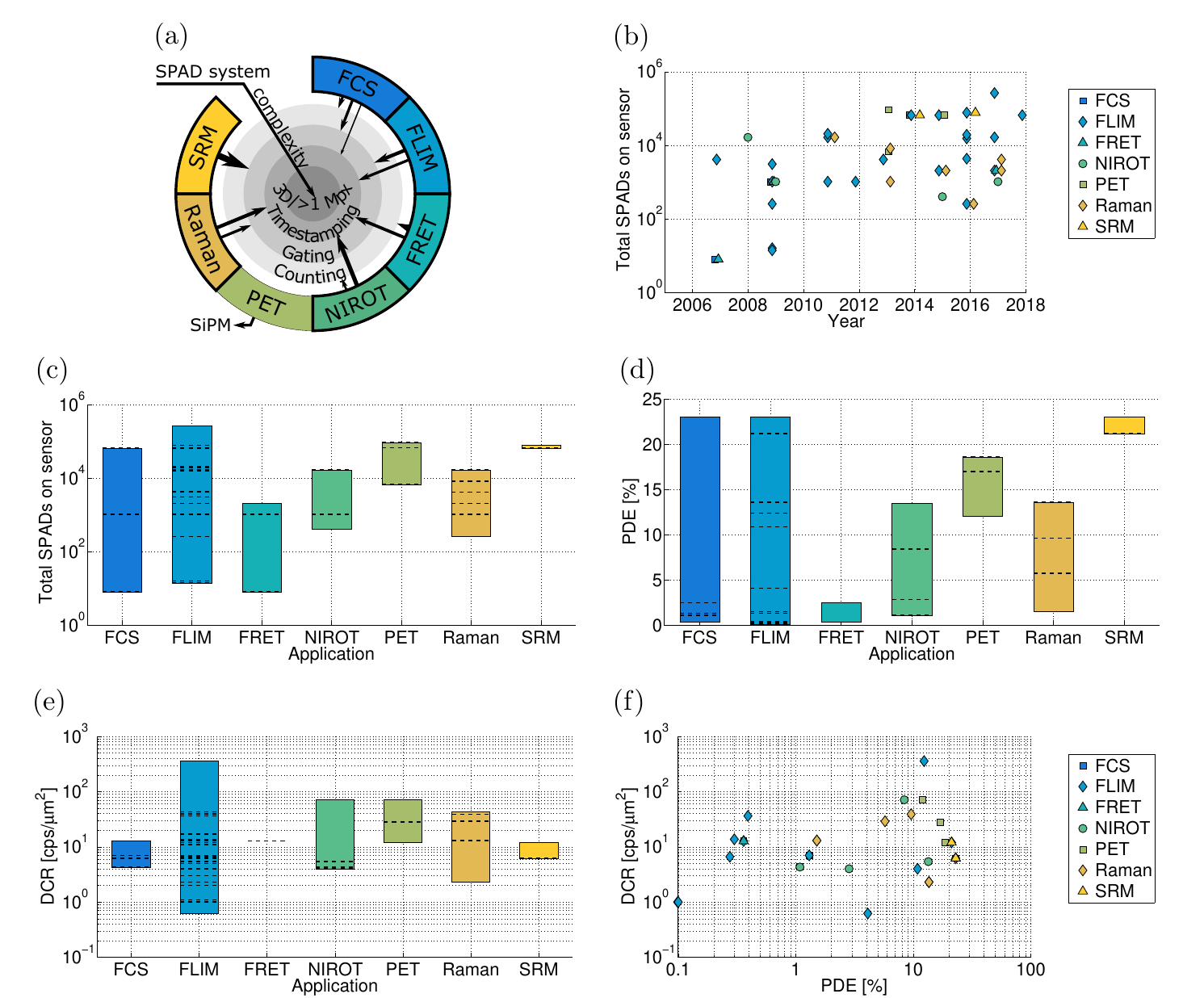}
		\caption{\label{fig:summary_SPADs}(a) Schematic overview of the SPAD-based system complexity, in terms of key functionalities (counting/gating/timestamping) vs. the main biophotonics applications. (b--f) Overview of representative SPAD sensor figures of merit as a function of the main target applications, based on data from \autoref{tab:summary_table}: (b) Total number of SPADs (corresponding to the effective spatial resolution in imaging arrays) vs. time; (c--e) Total number of SPADs, PDE and DCR per unit area grouped per application type (dashed lines: individual sensors, top/bottom of each box: maximum/minimum); (f) DCR per unit area vs. PDE.}
	\end{figure}

\autoref{fig:summary_SPADs}(a) provides a schematic overview of how the key functionalities of the SPAD-based imagers  reviewed in this paper are related to the main biophotonics applications to which they have been applied so far. The arrows pointing towards the center provide a qualitative indication of the relative importance of the different functionalities, whereby most applications have seen the use of at least two of them (e.g. timestamping and gating), excepted localization-based super-resolution microscopy (SRM), which has only needed photon counting so far. PET is a somewhat special case, relying on \emph{ad hoc} architectures as discussed in the first paragraph of the previous Section.

\autoref{fig:summary_SPADs}(b) shows the distribution of the total number of SPADs over time, based on the data from \autoref{tab:summary_table}, indicating a clear trend towards larger SPAD arrays over the past decade. (c--e) illustrate the distribution of some of the key figures of merit (total number of SPADs, PDE, and DCR per unit area), using again  \autoref{tab:summary_table} as a reference, grouped per main biophotonics application type. Each dashed line corresponds to an individual sensor. For the total number of SPADs and the PDE, higher values are preferred, whereas the contrary is obviously true for the DCR. The size of each box is representative of the spread of the figures of merit for that specific application; it is for example larger for FLIM, where the first developments have taken place early on, when the technology and designs were far from optimized. The box size also reflects the number of sensors which have been employed so far; from this perspective, it is smaller for SRM, which has only been addressed in the very recent past.

Finally, as discussed in the SPAD section the design of SPAD sensors entails numerous trade-offs, such as PDE vs. DCR shown in \autoref{fig:summary_SPADs}(f). This plot indicates a move towards higher PDE values, although this is not easy to achieve whilst still keeping low DCR (the ideal sensor would be placed in the bottom right corner of this plot).

 \section{Conclusions and outlook}\label{sec:conclusion} 

In this review, we have focused on SPAD imagers in standard CMOS technologies and their biophotonics applications. Individual SPAD pixels and small arrays have evolved from a scientific curiosity 15~years ago to a range of fully integrated devices, where the key challenges have been sensitivity, homogeneity, noise, reliability, and reproducibility. 

At device level, the performance gap between SPADs fabricated in custom technologies and in standard CMOS has been shrinking over time, whereby CMOS can leverage 60~years of experience and investments in scalable and high-yield technologies.  The corresponding improvements have been significant in terms of key parameters such as PDP and fill factor, with peak values in excess of 40--55\%, to DCR, drastically reduced to below 0.1~cps/$\mu$m$^{2}$ in the best SPADs, as well as the percentage of hot pixels (just a few percent in the best technologies). The spatial resolution has also increased to a quarter megapixel for the largest formats, with megapixel arrays on the horizon. This has called for a shrinkage of pixel size to well below 10~$\mu$m, with some groups targeting SPADs of just a few microns in deep-sub-micron technology nodes (45~nm and below) \cite{You2017}, not very far from their counterparts in CMOS imagers; this feat was unthinkable just a few years ago. The timing accuracy was already excellent (typically 50--100~ps, with the best SPADs in the 20--30~ps range), and has therefore evolved in a less spectacular fashion.

At the architectural level, standard CMOS SPAD arrays have a clear advantage over CCDs/\-sCMOS imagers, when parallel single-photon counting and/or timing is required; they can be coupled in a flexible way to different digital blocks for data acquisition and/or processing, in close sensor proximity, thanks to the natively digital SPAD data output. The absence of read-out noise is an important issue which is often neglected, all the more so when SPAD imagers are coupled to very high speed (about 100~kfps), ADC-less binary implementations, delivering continuous microsecond frames in real-time to capture fast transient phenomena. The capability of implementing integrated, parallel nanosecond gating is also a very appealing alternative to non-all-solid-state implementations.

It therefore comes to no surprise that FLIM has been explored first, together with multi-beam approaches that make full use of the sensors’ native parallelism. This was then followed by a host of other time-resolved biophotonics applications, all the way to disruptive scenarios such as quantum-based superresolution microscopy; the latter represents a good example of a sensor architecture which has been custom-designed to meet specific requirements, and for which efficient sensor alternatives are rare. It is however true that most SPAD imagers are still research prototypes, and only some are available commercially, e.g. from SMEs such as PhotonForce (UK) or MPD (Italy), usually as a spin-off of designs explored in academic environments. (Other commercially available SPAD arrays are usually derived from non-imaging SiPMs, e.g. the Philips Digital Photon Counting dSiPM for clinical PET, or the STMicroelectronics time-of-flight sensors \cite{Pellegrini2017}, aimed at ranging in the consumer market.)

This is partly due to the maturity of the sensors themselves and/or of the underlying technology, and partly to the fact that the overall sensitivity, in particular in the red, still lags behind those of CCDs/sCMOS imagers by 2--10$\times$. This can be an issue in applications where absolute sensitivity, rather than SNR, is a must, or where the illumination power needs to be kept low to avoid sample degradation. Certain applications do need a very high timing resolution, typical of single, highly optimized SPADs, and can tolerate longer acquisition times due to scanning. Finally, in certain cases SPAD designers need to face competition from other single-photon or established technologies.

In the future, we foresee that academic and research establishments will continue pushing the state-of-the-art in terms of the key figures of merit, targeting large format and high performance CMOS SPAD arrays, with industrial applications starting to appear in the medium term. Larger industrial conglomerates are more likely to emphasize high volume applications in the mobile/consumer areas, e.g covering automotive, point-of-care, and Internet of Things (IoT), possibly via smaller, dedicated units; these foundries might also propose SPADs as IP blocks in a not too distant future. The resulting consumer oriented developments are likely to follow trends similar to the ones which characterized high end smartphone imagers (i.e. towards chip stacking and 3D ICs); they might impact in a positive way niche markets in general, and time-resolved applications of interest to the biophotonics community in particular.

\bibliographystyle{naturemag}
\afterpage{
\clearpage
    \pagestyle{empty}
	\newgeometry{margin=1.5cm}
    \begin{landscape}
        \centering
        \footnotesize
        \captionsetup{type=table}
        \caption{\label{tab:summary_table}Overview of main SPAD sensors and imagers targeted at biophotonics applications, in chronological order, as published over the past fifteen years. All values and operating modes as reported in literature.
        \\Definitions: \emph{PDE}: SPAD photon detection probability at nominal excess bias voltage, multiplied by the pixel fill factor; \emph{DCR}: median dark count rate per SPAD unit area, for the same excess bias voltage as the PDE.
        \\Operating mode definitions: \emph{TCSPC}: time-correlated single-photon counting; \emph{Gating}: use of one or multiple moving gates; \emph{Majority time voting} generation of a time stamp per event (on the first arrived photon in a pixel, in the simplest case), only if a certain photon count is reached.}
        \rowcolors{2}{gray!14}{white}
\begin{longtable}{|P{2.6cm}c|C{1.4cm}C{1.25cm}C{1.25cm}C{1.35cm}C{1.25cm}C{1.2cm}P{1.4cm}P{2.4cm}P{2.4cm}P{1.4cm}|}
\hline
\rowcolor{gray!30}
Sensor \& Architecture    & Year                   & SPAD array  & Technology [nm] &	SPAD diameter$_\text{eq.}$ [$\mu$m] & Fill factor [\%]       & PDE$_\text{top}$ [\%]  & DCR [cps/$\mu$m$^2$]  & Timing technique       & Sensor specifications  & System features        & Application \\

\hline

    First CMOS SPAD array \cite{Rochas2003}    & 2003                 & 8$\times$4           & -                    & 6.4                  & $<$ 1                  & 0.2                  & 1.6                  & -                    & -                    &                      & - \\
	
	\hdashline
	
    Rech                 \cite{Rech2007, Rech2009, Michalet2010, Ingargiola2012, Ingargiola2013, Michalet2013, Michalet2014} & 2007                 & 8$\times$1           & -                    & 50.0                 & 5                    & 2.5                  & 1.0                  & -                    & -                    &                      & FRET / FCS \\
    Schwartz             \cite{Schwartz2007, Schwartz2008} & 2007                 & 64$\times$64         & 350                 & 4.1                  & $<$ 1                  & 0.1                  & 71.0                 & TCSPC + Gating       & In-pixel TDC         & 4096 in-pixel 350~ps 10b TDCs & FLIM \\
    Niclass (LASP)       \cite{Niclass2008, Pavia2014II, Pavia2014} & 2008                 & 128$\times$128       & 350                 & 7.0                  & 6, $\times2-8^\text{ml}$ & 2.1, $\times2-8^\text{ml}$                  & 17.0                 & TCSPC                & Column-based TDCs    & 32 column 98~ps 10b TDCs & NIROT \\
    Boiko (G$^{(2)}$)    \cite{Boiko2009, Boiko2009II} & 2009                 & 4$\times$4           & 350                 & 3.5                  & -                    & -                    & 1.0                  & -                    & -                    &                      &  \\
    Niclass (FluoCAM)    \cite{Niclass2009, Stegehuis2015, Homulle2016_FluoCAM} & 2009                 & 60$\times$48         & 350                 & 8.6                    & $<$ 1                  & 0.1                  & 7.0                  & Gating (2$\times$)   & 2 in-pixel 8b counters & 5~ns gate, 12~ps steps & FLIM \\
    Guerrieri            \cite{Guerrieri2009, Guerrieri2010, Colyer2011, Vitali2014} & 2009                 & 32$\times$32         & 350                 & 20.0                 & 3.1                  & 1.3                  & 12.7                 & Gating               & In-pixel 8b counter  &                      & FLIM / FCS \\
    MEGAFRAME32         \cite{Richardson2009, Li2010, Li2011, Li2012, Gersbach2010, Gersbach2012, Kloster2013, Krstajic2015II, Poland2013, Poland2014, Poland2015, Poland2016} & 2009                 & 32$\times$32         & 130                 & 5.6                    & 1                    & 0.4                  & 4.0                  & TCSPC                & In-pixel TDC         & 1024 in-pixel 50~ps 10b TDCs & FLIM / FCS / FRET \\
    Pancheri             \cite{Pancheri2009}  & 2009                 & 64$\times$4          & 350                 & 17.6                  & 34                   & 10.9                 & 4.3                  & Gating (4$\times$)   & 4 in-pixel 8b counters & 4 SPADs = 1 pixel    & FLIM \\
    Carrara (RadHard2)   \cite{Carrara2009, Stuker2010, Stuker2011, Singh2013} & 2009                 & 32$\times$32         & 350                 & 6.0                  & 3.1                  & 1.1                  & 5.0                  & -                    & In-pixel 1b counter  &                      & FCS / NIROT \\
    Stoppa               \cite{Stoppa2009}    & 2009                 & 7$\times$2           & 350                 & -                    & -                    & -                    & 13.0                 & Gating               & In-pixel 17b counter &                      & FLIM \\
    Maruyama             \cite{Maruyama2011, Blacksberg2011} & 2011                 & 128$\times$128       & 350                 & 6.0                    & 4.5, $\times1.6^\text{ml}$ & 0.9, $\times1.6^\text{ml}$                  & 6.6                  & Gating               & In-pixel 1b counter  &                      & FLIM / Raman \\
    MEGAFRAME128        \cite{Veerappan2011, Veerappan2011ES, Arlt2011} & 2011                 & 160$\times$128       & 130                 & 5.6                    & 1                    & 0.3                  & 2.0                  & TCSPC                & In-pixel TDC         & 20480 in-pixel 55~ps 10b TDCs & FLIM \\
    Pancheri             \cite{Pancheri2011, Pancheri2012, Pancheri2013} & 2011                 & 32$\times$32         & 350                 & 12.9                 & 20.3                 & -                    & 5.4                  & Gating               & In-pixel analog counter & 1.9~ns gate          & FLIM \\
	
	\hline
\rowcolor{gray!30}
Sensor \& Architecture    & Year                   & SPAD array  & Technology [nm] &	SPAD diameter$_\text{eq.}$ [$\mu$m] & Fill factor [\%]       & PDE$_\text{top}$ [\%]  & DCR [cps/$\mu$m$^2$]  & Timing technique       & Sensor specifications  & System features        & Application \\

\hline
	
    Durini (BackSPAD)    \cite{Durini2012, Zou2014} & 2012                 & 32$\times$32         & 350$^\text{3D}$     & 94.4                 & 75.4                 & -                    & 39.7                 & -                    & In-pixel counters    & Preliminary          &  \\
    Tyndall              \cite{Tyndall2012I, Tyndall2012II, Arlt2013} & 2012                 & 32$\times$32         & 130                 & 8.0                  & 10                   & -                    & 13.7                 & TCSPC                & Per group TDC        & 16 52~ps 16b TDCs, mini-SiPM    & FLIM \\
    Field                \cite{Field2013, Field2014} & 2013                 & 64$\times$64         & 130                 & 5.0                  & $<$ 1                  & 0.3                  & 28.0                 & TCSPC                & Column-based TDCs    & 4096 column 62.5~ps 10b TDCs & FLIM \\
    Mandai               \cite{Mandai2013}    & 2013                 & 416$\times$4$\times$4 & 350                 & 32.6                    & 55.6                 & 17.0                 & 39.0                 & Majority time voting & Column-based per group TDC + in-pixel 1b counter & 192 column 44~ps 17b TDCs & PET \\

    Maruyama             \cite{Maruyama2013, Maruyama2014} & 2013                 & 1024$\times$8        & 350                 & 18.0                    & 44.3                 & 9.6                  & 29.0                 & Gating               & In-pixel 1b counter  & 0.7~ns gate, 250~ps steps & Raman \\
    Nissinen             \cite{Nissinen2015, Nissinen2015II, Nissinen2013, Kostamovaara2013} & 2013                 & 128$\times$8         & 350                 & 9.7                 & 23                   & 5.8                  & 71.0                 & Gating (4$\times$)   & 4-pixel gate comparators & 4 SPADs = 1 pixel    & Raman \\
    Walker (SPADnet 1)   \cite{Walker2013, Braga2014, Unternahrer2016, Gasparini2017} & 2013                 & 720$\times$16$\times$8 & 130                 & 16.3                 & 42.9                 & 12.0                 & 6.2                  & Majority time voting & In-pixel TDC + 7b counter & 128 in-pixel 64~ps 12b TDCs + histogramming & PET \\
    Burri (SwissSPAD)    \cite{Burri2014_SwissSPAD, Antolovic2016, Antolovic2016II, Antolovic2016III, Antolovic2017, Wargocki2017} & 2014                 & 512$\times$128       & 350                 & 6.0                  & 5, $\times8-12^\text{ml}$ & 2.3, $\times8-12^\text{ml}$                  & 12.0                 & Gating               & In-pixel 1b counter  & 4~ns gate, 20~ps steps & FLIM / FCS / SRM \\
    Carimatto            \cite{Carimatto2015} & 2015                 & 416$\times$18$\times$9 & 350                 & 33.0                    & 57                   & 18.6                 & 43.0                 & Majority time voting & Column-based per group TDC + in-pixel 1b counter & 432 column 48~ps 17b TDCs & PET \\
    Krstaji{\'c}         \cite{Krstajic2015, Ehrlich2017} & 2015                 & 256$\times$8         & 130                 & 18.2                 & 43.7                 & -                    & 5.4                  & TCSPC + Gating       & Per-pixel TDC + histograms & 512 per-pixel 40~ps TDCs + histogramming & FLIM / Raman \\
    Parmesan             \cite{Parmesan2015}  & 2015                 & 256$\times$256       & 130                 & 4.2                  & 19.6                 & -                    & 4.0                  & TCSPC                & TAC pixels           & External 14b ADC     & FLIM \\
    Mata Pavia (3DAPS)   \cite{Pavia2015, Pavia2015_thesis} & 2015                 & 400$\times$1         & 130$^\text{3D}$     & 6.0                    & 23.3                 & 2.8                  & 357.0                & TCSPC                & In-pixel TDC         & 3D stacked, 50~ps 12b TDCs & NIROT \\
    Abbas                \cite{Abbas2016}     & 2016                 & 128$\times$120       & 65$^\text{3D}$    & 5.9                  & 45                   & 12.4                 & 36.2                 & Gating               & In-pixel 12b counter &                      &  \\
    Lee                  \cite{Lee2016}       & 2016                 & 72$\times$60         & 180                 & 15.0                 & 14.4                 & 0.4                  & 2.3                  & Gating               & In-pixel 10b counter & 10~ns gate, 72~ps steps & FLIM \\
	
	\hline
\rowcolor{gray!30}
Sensor \& Architecture    & Year                   & SPAD array  & Technology [nm] &	SPAD diameter$_\text{eq.}$ [$\mu$m] & Fill factor [\%]       & PDE$_\text{top}$ [\%]  & DCR [cps/$\mu$m$^2$]  & Timing technique       & Sensor specifications  & System features        & Application \\

\hline
	
    Burri (LinoSPAD)     \cite{Burri2016_LinoSPAD, Burri2017} & 2016                 & 256$\times$1         & 350                 & 17.1                    & 40                   & 13.6                 & 11.0                 & TCSPC (External)     & -                    & 64 FPGA-based 25~ps TDCs & FLIM / Raman \\
	
    Perenzoni            \cite{Perenzoni2016} & 2016                 & 160$\times$120       & 350                 & 7.8                    & 21                   & -                    & 12.0                 & Gating               & Column analog counter & 10~ns gate, 194~ps steps & FLIM \\
    Dutton (SPCIMAGER)   \cite{Dutton2016, Gyongy2016, Gyongy2018} & 2016                 & 320$\times$240       & 130                 & 4.7                    & 26.8, $\times1.8-2^\text{ml}$ & 10.6, $\times1.8-2^\text{ml}$                 & 3.0                  & Gating               & In-pixel analog counter &                      & FLIM / SRM \\
    Erdogan              \cite{Erdogan2017}   & 2017                 & 1024$\times$16       & 130                 & 18.8                    & 49.3                 & -                    & -                    & TCSPC + Gating       & Per-pixel TDC + histograms & 512 per-pixel 50~ps TDCs + histogramming & FLIM \\
    Holma                \cite{Holma2017, Nissinen2018} & 2017                 & 256$\times$16        & 350                 & 18.0                 & 26                   & -                    & -                    & TCSPC                & Shared TDCs    & Two 52~ps 3b TDCs    & Raman \\
    Kufcs{\'a}k              \cite{Kufcsak2017}   & 2017                 & 256$\times$8         & 130                 & 18.2                 & 43.7                 & -                    & 5.4                  & TCSPC + Gating       & Per-pixel TDC + histograms & Improvement of \cite{Krstajic2015} & FLIM / FRET / Raman \\
    Lindner (Piccolo) \cite{Lindner2017, Lindner2018, Kalyanov2018} & 2017                 & 32$\times$32         & 180                 & 17.0                 & 28                   & 13.4                 & 0.6                  & TCSPC                & Column-based TDCs, dynamic reallocation    & 128 column 49~ps TDCs & NIROT \\
    Ulku (SwissSPAD2)    \cite{Ulku2017}      & 2017                 & 512$\times$512       & 180                 & 6.0                 & 10.5                 & 5.2                  & 0.3                  & Gating               & In-pixel 1b counter  & 5~ns gate            & FLIM \\
    Gyongy               \cite{Gyongy2018TED} & 2018                 & 256$\times$256       & 130                 & 14.1                 & 61                   & -                    & 51.0                 & Gating               & In-pixel 1b counter  &                      & FLIM \\

\hline
\end{longtable}

         \vspace{-10pt}
        $^\text{ml}$: use of microlenses -- the quoted native PDE / fill factor need to be multiplied by a concentration factor.\\
		$^\text{3D}$: use of a 3D integration technology (or trial). 
    \end{landscape}
    \clearpage}

\end{document}